\documentclass[twoside,twocolumn,9pt]{article}
\usepackage{extsizes}
\usepackage[super,sort&compress,comma]{natbib} 
\usepackage[version=3]{mhchem}
\usepackage[left=1.5cm, right=1.5cm, top=1.785cm, bottom=2.0cm]{geometry}
\usepackage{balance}
\usepackage{mathptmx}
\usepackage{sectsty}
\usepackage{graphicx} 
\usepackage{lastpage}
\usepackage[format=plain,justification=justified,singlelinecheck=false,font={stretch=1.125,small,sf},labelfont=bf,labelsep=space]{caption}
\usepackage{float}
\usepackage{fancyhdr}
\usepackage{fnpos}
\usepackage[english]{babel}
\addto{\captionsenglish}{%
  
}
\usepackage{array}
\usepackage{droidsans}
\usepackage{charter}
\usepackage[T1]{fontenc}
\usepackage[usenames,dvipsnames]{xcolor}
\usepackage{setspace}
\usepackage[compact]{titlesec}
\usepackage{hyperref}

\definecolor{cream}{RGB}{222,217,201}

\begin{document}

\pagestyle{fancy}
\thispagestyle{plain}
\fancypagestyle{plain}{
\renewcommand{\headrulewidth}{0pt}
}

\makeFNbottom
\makeatletter
\renewcommand\LARGE{\@setfontsize\LARGE{15pt}{17}}
\renewcommand\Large{\@setfontsize\Large{12pt}{14}}
\renewcommand\large{\@setfontsize\large{10pt}{12}}
\renewcommand\footnotesize{\@setfontsize\footnotesize{7pt}{10}}
\makeatother

\renewcommand{\thefootnote}{\fnsymbol{footnote}}
\renewcommand\footnoterule{\vspace*{1pt}%
\color{cream}\hrule width 3.5in height 0.4pt \color{black}\vspace*{5pt}} 
\setcounter{secnumdepth}{5}

\makeatletter 
\renewcommand\@biblabel[1]{#1}            
\renewcommand\@makefntext[1]%
{\noindent\makebox[0pt][r]{\@thefnmark\,}#1}
\makeatother 
\renewcommand{\figurename}{\small{Fig.}~}
\sectionfont{\sffamily\Large}
\subsectionfont{\normalsize}
\subsubsectionfont{\bf}
\setstretch{1.125} 
\setlength{\skip\footins}{0.8cm}
\setlength{\footnotesep}{0.25cm}
\setlength{\jot}{10pt}
\titlespacing*{\section}{0pt}{4pt}{4pt}
\titlespacing*{\subsection}{0pt}{15pt}{1pt}

\fancyfoot{}
\fancyfoot[LO,RE]{\vspace{-7.1pt}\includegraphics[height=9pt]{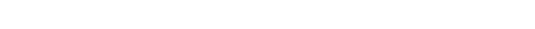}}
\fancyfoot[CO]{\vspace{-7.1pt}\hspace{11.9cm}\includegraphics{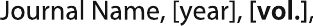}}
\fancyfoot[CE]{\vspace{-7.2pt}\hspace{-13.2cm}\includegraphics{head_foot/RF}}
\fancyfoot[RO]{\footnotesize{\sffamily{1--\pageref{LastPage} ~\textbar  \hspace{2pt}\thepage}}}
\fancyfoot[LE]{\footnotesize{\sffamily{\thepage~\textbar\hspace{4.65cm} 1--\pageref{LastPage}}}}
\fancyhead{}
\renewcommand{\headrulewidth}{0pt} 
\renewcommand{\footrulewidth}{0pt}
\setlength{\arrayrulewidth}{1pt}
\setlength{\columnsep}{6.5mm}
\setlength\bibsep{1pt}

\makeatletter 
\newlength{\figrulesep} 
\setlength{\figrulesep}{0.5\textfloatsep} 

\newcommand{\topfigrule}{\vspace*{-1pt}%
\noindent{\color{cream}\rule[-\figrulesep]{\columnwidth}{1.5pt}} }

\newcommand{\botfigrule}{\vspace*{-2pt}%
\noindent{\color{cream}\rule[\figrulesep]{\columnwidth}{1.5pt}} }

\newcommand{\dblfigrule}{\vspace*{-1pt}%
\noindent{\color{cream}\rule[-\figrulesep]{\textwidth}{1.5pt}} }

\makeatother

\twocolumn[
  \begin{@twocolumnfalse}
{\includegraphics[height=30pt]{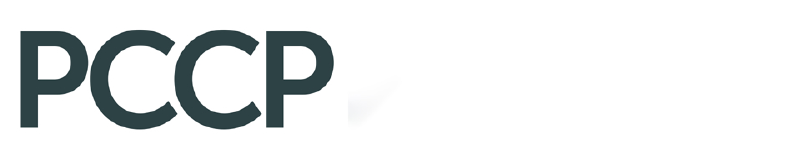}\hfill\raisebox{0pt}[0pt][0pt]{\includegraphics[height=55pt]{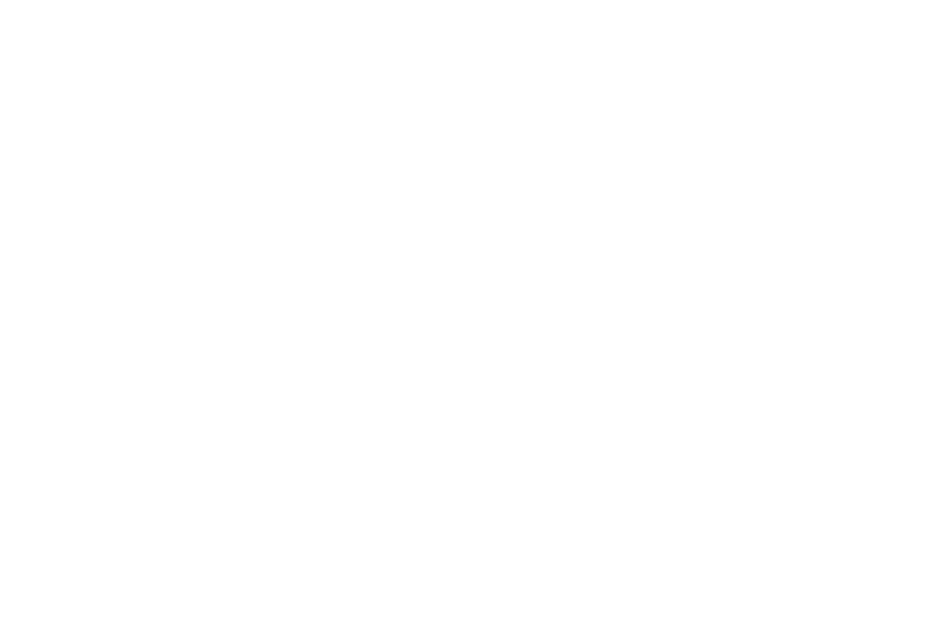}}\\[1ex]
\includegraphics[width=18.5cm]{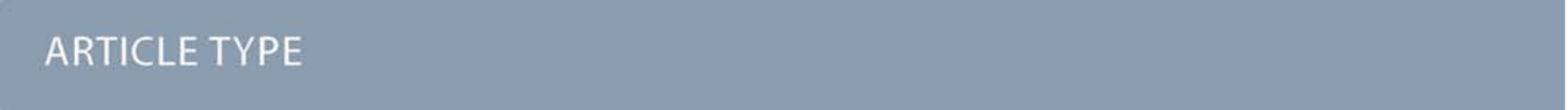}}\par
\vspace{1em}
\sffamily
\begin{tabular}{m{4.5cm} p{13.5cm} }

\includegraphics{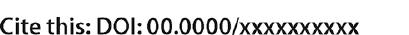} & \noindent\LARGE{\textbf{Is the Doped \ce{MoS2} Basal Plane an Efficient Hydrogen Evolution Catalyst? Calculations of Voltage-Dependent Activation Energy
}
} \\
\vspace{0.3cm} & \vspace{0.3cm} \\

& \noindent\large{Sander \O. Hanslin,\textit{$^{a,b}$} Hannes J\'onsson,\textit{$^{b,c}$} and Jaakko Akola\textit{$^{a,d,\ddag}$}} \\

\includegraphics{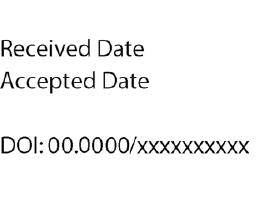} & \noindent\normalsize{Transition metal dichalcogenides are cheap and earth-abundant candidates for the replacement of precious metals as catalyst materials. Experimental measurements of the hydrogen evolution reaction (HER), for example, have demonstrated significant electrocatalytic activity of \ce{MoS2} but there is large variation depending on preparation method.
In order to gain information about the mechanism and active sites for HER, we have carried out calculations of the reaction and activation energy for HER at the transition metal doped basal plane of \ce{MoS2} under electrochemical conditions, {\it i.e.} including applied electrode potential and solvent effects.
The calculations are based on identifying the relevant saddle points on the energy surface obtained from density functional theory within the generalized gradient approximation, and the information on energetics is used to construct voltage-dependent volcano plots. Doping with 3d-metal atoms as well as \ce{Pt} is found to enhance hydrogen adsorption onto the basal plane by introducing electronic states within the band gap, and in some cases (\ce{Co}, \ce{Ni}, \ce{Cu}, \ce{Pt}) significant local symmetry breaking.
The Volmer-Heyrovsky mechanism is found to be most likely and the associated energetics show considerable dopant and voltage-dependence. While the binding free energy of hydrogen can be tuned to be seemingly favorable for HER, the calculated activation energy turns out to be significant, at least $0.7$~eV at a voltage of $-0.5$~V vs. SHE, indicating low catalytic activity of the doped basal plane.
This suggests that other sites are responsible for the experimental activity, possibly edges or basal plane defects.
}

\end{tabular}

 \end{@twocolumnfalse} \vspace{0.6cm}

  ]

\renewcommand*\rmdefault{bch}\normalfont\upshape
\rmfamily
\section*{}
\vspace{-1cm}


\footnotetext{\textit{$^{a}$~Department of Physics, Norwegian University of Science and Technology, NO-7491 Trondheim, Norway }}
\footnotetext{\textit{$^{b}$~Faculty of Physical Sciences and Science Institute, University of Iceland, IS-107 Reykjav{\'\i}k, Iceland}}
\footnotetext{\textit{$^{c}$~Applied Physics Department, Aalto University,  FI-00076 Aalto, Finland}}
\footnotetext{\textit{$^{d}$~Computational Physics Laboratory, Tampere University, FI-33101 Tampere, Finland}}

\footnotetext{\dag~Electronic Supplementary Information (ESI) available. See DOI: 10.1039/cXCP00000x/}
\footnotetext{\ddag~E-mail: jaakko.akola@ntnu.no}

\section{Introduction}

\noindent 
Hydrogen is considered as one of the most promising means of 
future storage of 
renewable energy from intermittent sources \cite{DAWOOD20203847}.
Gaseous hydrogen can be produced through electrolysis of water
in a potentially cheap and sustainable way of converting 
electrical energy from renewable sources into chemical energy. 
The efficiency and cost of this process largely depends on 
the catalyst material. Currently, the process relies heavily on the high activity of platinum-group metals (PGMs), and particularly platinum itself \cite{HOLLADAY2009244}. The limited availability of these precious metals in Earth's crust and socioeconomic issues in the mining countries pose problems for long-term sustainability and calls for the development of new catalyst materials as well as better understanding of the fundamental mechanisms of the hydrogen evolution reaction (HER) and electrocatalysis in general. Among the emerging candidates in the search of sustainable replacements, various metal alloys and transition metal compounds have been found \cite{C4EE01760A}. Transition metal dichalcogenides have been shown to exhibit promising properties for HER, and 
\ce{MoS2} has,
in particular,
been widely investigated both experimentally and theoretically \cite{doi:10.1021/ja0504690,B803857K,doi:10.1021/cs300451q,Kibsgaard2012}. 

An important step in characterizing and designing new electrocatalysts is to identify the atomic sites that exhibit 
high activity.
While the edge sites of pristine 2H-\ce{MoS2} have been shown to be catalytically active, the basal plane 
is inert in its 
pure
form \cite{doi:10.1021/ja0504690,doi:10.1126/science.1141483}. \ce{MoS2} has been synthesized in a wide range of morphologies \cite{doi:10.1021/nl400258t} with a focus on increasing the 
abundance
of edge sites. Further, the basal plane can be activated by introducing defects such as \ce{S}-vacancies \cite{Li2016}, phase boundaries \cite{C9CY01901D} and impurities \cite{C5EE00751H}.
As such, the full picture of \ce{MoS2} activity is
quite complex and a basic understanding of the features in atomic and electronic structure that enhance HER is needed to optimize performance.

Experiments have indicated that transition metal doping enhances the overall activity \cite{Wang2015,C2SC20539D,doi:10.1021/acscatal.6b01274}, but 
can also have
detrimental effects and conflicting results have been 
reported
\cite{C8SC01114A,C9NR10702A}, illustrating that the result strongly depends on the system specifics, notably morphology, the nature and level of doping, and 
the 
experimental techniques
applied.
This indicates that the manifested activity 
relies on the interplay between several factors
and that theoretical studies can therefore be helpful for identifying the 
contributing ones.

In the present study, we investigate the effect of transition metal doping on the electrocatalytic activity of the \ce{MoS2} basal plane.
Theoretical studies have shown that hydrogen adsorption onto the basal plane is enhanced by transition metal doping \cite{Hakala2017}, and in the following we will assess whether this corresponds to higher activity (reduced reaction barriers) under electrochemical conditions for the whole sequence of 3d-metals as well as platinum. First we investigate the doped material itself and hydrogen adsorption in the gas phase. Then we move on to model the electrochemical reactions involved in hydrogen evolution. These are more challenging than calculations of gas/surface reactions because:
i) the reaction occurs in the presence of an electrolyte and ii) the reaction occurs at a fixed electrode potential. These challenges must be overcome to provide a realistic model
for electrochemical reactions, as further discussed in the following section. 
\section{Methods}

\subsection{Calculations of Activation Energy}

All results were obtained from spin-polarized density-functional theory (DFT) calculations within the generalized gradient approximation (GGA). 
The revised Perdew-Burke-Ernzerhof (rPBE) exchange-correlation functional by Hammer {\it et al.} \cite{PhysRevB.59.7413} was used because of its 
improved results for adsorption energy. In addition, van der Waals interactions were accounted for by the zero-damping D3 parameters
\cite{doi:10.1063/1.3382344} as this provides an improved description of the interlayer distance of multilayer \ce{MoS2}. A cutoff of $400$~eV was used for the plane wave kinetic energy
in the representation of the valence electrons, and the
projector augmented wave (PAW) \cite{PhysRevB.50.17953} approach was used to represent the effect of inner electrons.
For all transition metals, the outermost $s$- and $d$-electrons were treated as valence electrons. 
For \ce{O} and \ce{S}, the 2s$^2$2p$^4$ and 3s$^2$3p$^4$ electrons were treated as valence, respectively. Test calculations including also $3s$- and $3p$-electrons for the early 3d-metals did not indicate any discrepancy in adsorption energies nor in the local electronic structure at the adsorption sites.
 As GGA tends to excessively delocalize the wave function due to self-interaction, the description of d-states was compared to that of the Hubbard U approach \cite{doi.org/10.1002/qua.24521} and a hybrid functional (PBE0 \cite{doi:10.1063/1.472933, doi:10.1063/1.478522}), see Figures S4 and S5$^\dag$.
The Vienna Ab Initio Simulation Package (VASP) was used in the DFT simulations \cite{PhysRevB.59.1758}.

The geometry optimization of bulk \ce{MoS2} was performed 
using
the primitive unit cell of 2H-\ce{MoS2}, with the Brillouin zone sampled by a Monkhorst-Pack (MP) grid of dimension $9\times9\times5$. For calculations on doped mono- and bilayers
we consider $5\times5\times1$ supercells (75 atoms), with $3\times3\times1$ MP grids. For accurate density of states (DOS) calculations, $11\times11\times1$ MP grids were used. A vacuum layer of ca. $14$~{\AA} was introduced to decouple the periodic images of 
the slab.

The activation energy for the various elementary steps in the electrochemical reaction was calculated by first finding an approximate minimum energy path for the transition using the climbing-image nudged elastic band (CI-NEB) \cite{doi:10.1063/1.1329672} method, followed by calculations with a tighter convergence as the saddle point on the energy surface corresponding to a given applied voltage is found using the minimum mode following (MMF)  \cite{doi:10.1063/1.480097} method. The tolerance for force convergence in saddle point searches was set at $0.05$~eV/{\AA} while the tolerance in minimization calculations was $0.02$~eV/{\AA}. 

\subsection{Reaction Mechanism
}
The hydrogen evolution reaction involves the adsorption of \ce{H+} from solution onto the catalyst surface and subsequent desorption of gaseous \ce{H2}. This process can be described in terms of three steps:
\begin{align}
&\ce{H+ + e- -> H^*}\\
&\ce{H+ + H^* + e- -> H2}\\
&\ce{2H^* -> H2}
\end{align}
where \ce{H^*} indicates hydrogen bound to a surface site.
Adsorption occurs through the Volmer (1) mechanism, and desorption through either the Heyrovsky (2) or Tafel (3) mechanisms. Whether the evolution proceeds through the Tafel or Heyrovsky mechanism (or a combination of both) depends on the kinetics of these in the given system. 

The most widely used descriptor for the HER efficiency of a material is the free energy of hydrogen adsorption $\Delta G_{\mathrm{H}} $ on its surface. In accordance with 
the Sabatier principle, a value of $\Delta G_{\mathrm{H}}\approx 0$ has been shown to correlate with high exchange currents \cite{TRASATTI1972163,N_rskov_2005}. In the gas phase, we define the $n$-th hydrogen adsorption energy as 
\begin{equation}
    \Delta E_{n\ce{H}} = E_{\ce{MoS2}+n\ce{H}} - E_{\ce{MoS2}+(n-1)\ce{H}} - \frac{n}{2} E_{\ce{H2}},
\end{equation}
Furthermore, the Gibbs free energy is then given in terms of this energy as
\begin{equation}
    \Delta G_{\ce{H}} = \Delta E_{\ce{H}} + \Delta E_{\mathrm{ZPE}} - T\Delta S_{\ce{H}}.
\end{equation}
We can approximate the entropic term as $\Delta S_{\ce{H}} = S_{\ce{H}^*} - \frac{1}{2}S^0_{\ce{H2}} \approx - \frac{1}{2}S^0_{\ce{H2}}$, where we neglect the configurational and vibrational entropy of the adsorbed state. At $298$~K, the entropy contribution is about $20$~meV. The zero-point energy $E_\mathrm{ZPE}$ is calculated individually for each system, and compared to the reference value of \ce{H2} (vib. frequency of $\approx 4400$~cm$^{-1}$), so that $\Delta E_\mathrm{ZPE} = E_\mathrm{ZPE}^{\mathrm{H}^*}-\frac{1}{2} E_\mathrm{ZPE}^{\mathrm{\ce{H2}}}$. 
We assume that the zero-point energy does not vary considerably for different hydrogen coverages.

Under electrochemical conditions, the chemical potential of a \ce{H+}-\ce{e-} pair in solution is given in terms of the electrode potential $U (\mathrm{V\,vs.\,SHE})$ and $\mathrm{pH}$ \cite{doi:10.1021/jp047349j} as
\begin{equation}
\mu_{\ce{H+}} + \mu_{\ce{e-}} = \frac{1}{2}\mu_{\ce{H2}} - e U + k_\mathrm{B}T \ln{a_{\mathrm{H}^+}},
\label{mu_H}
\end{equation}
where $k_\mathrm{B}$ is the Boltzmann constant, $T$ is the temperature and $a_{\mathrm{H}^+}$ is the activity of protons which is related to pH as $\mathrm{pH} = -\log{a_{\mathrm{H}^+}}$. 
Since we consider acidic solutions ($\mathrm{pH} \to 0$ corresponding to 1M [\ce{H+}]), the final term is negligible, and the chemical potential is essentially linearly modulated by the electrode potential.

\subsection{Solvent Model}

One of the challenges of modelling electrochemical reactions is to ensure an accurate description of the solvent. Explicit description of the solvent is computationally 
intensive
and the solvent atoms introduce a large number of degrees of freedom which complicates the process of finding 
energy minima
and saddle-point structures. In this work, we employ an implicit solvent model through the implementation in VASPsol \cite{doi:10.1063/1.4865107,doi:10.1063/1.5132354}, where the solvent (here water) is treated as a polarizable continuum. In this theory, the dielectric permittivity is spatially modulated between the extreme values of 1 (in vacuum) and $\epsilon_b$ (in bulk solvent) by a shape function $\zeta(\mathbf{r})$ as $\epsilon_r(\mathbf{r}) = 1 + (\epsilon_b - 1)\zeta(\mathbf{r})$.
The modulation defines regions of solute and solvent, depending on the local electron density $n(\mathbf{r})$ in terms of the complementary error function. The free energy is minimized by equating the variation with respect to both the electron density and the electrostatic potential to zero. The former leads to additional terms in the local potential of the Kohn-Sham equations, as described in detail in Ref. \citenum{doi:10.1063/1.5132354}. From the latter we obtain the generalized Poisson-Boltzmann equation, which describes the distribution of counter-ions in solution. 

The implicit solvent model allows us to largely omit explicit \ce{H2O} in the calculations, but to model a realistic stability and reactivity of the solvated \ce{H+}, some water molecules are still needed. An often used model is the hexagonal ice bilayer. In most systems, however, the periodicity of the system of interest is not commensurate with the typical hexagonal water structure. We therefore employ a cluster model based on the Eigen cation (\ce{H9O4+}), where a proton is effectively shared between four water molecules. For such a water model it follows that, since the ion is not externally restricted by hydrogen bonds, we expect to obtain a lower estimate of the activation energy compared to constrained structures such as hexagonal ice. 

Test calculations were performed to assess the size effect of  the water cluster.
Calculations with Hydronium (\ce{H3O+}) and Zundel (\ce{H5O2+}) ions showed respective discrepancies of at most $\sim0.24$~eV and $\sim0.04$~eV in the (Volmer/Heyrovsky) activation and reaction energies, with respect to the energies obtained with the Eigen cation (see Figure S1$^\dag$). This indicates reasonable convergence with increasing cluster size. 
We point out that
calculations using only a single hydronium ion in the absence of an implicit solvent yield 
very different results for
both the Volmer and Heyrovsky case. The 
reaction energies were decreased by roughly $1.1$~eV and $2.7$~eV, respectively, meaning that a single hydronium ion is far too unstable in the gas phase to provide reliable results. For the Eigen cation, the effect is smaller but still significant; the reaction energies are reduced by roughly $0.3$~eV and $0.9$~eV, respectively, in the absence of implicit solvent. Thus the implicit solvent is essential for the description of protonated water in combination with a cluster model.
\subsection{Applied Voltage}

\begin{figure}[!h]
    \centering
    \includegraphics[width=1.0\linewidth]{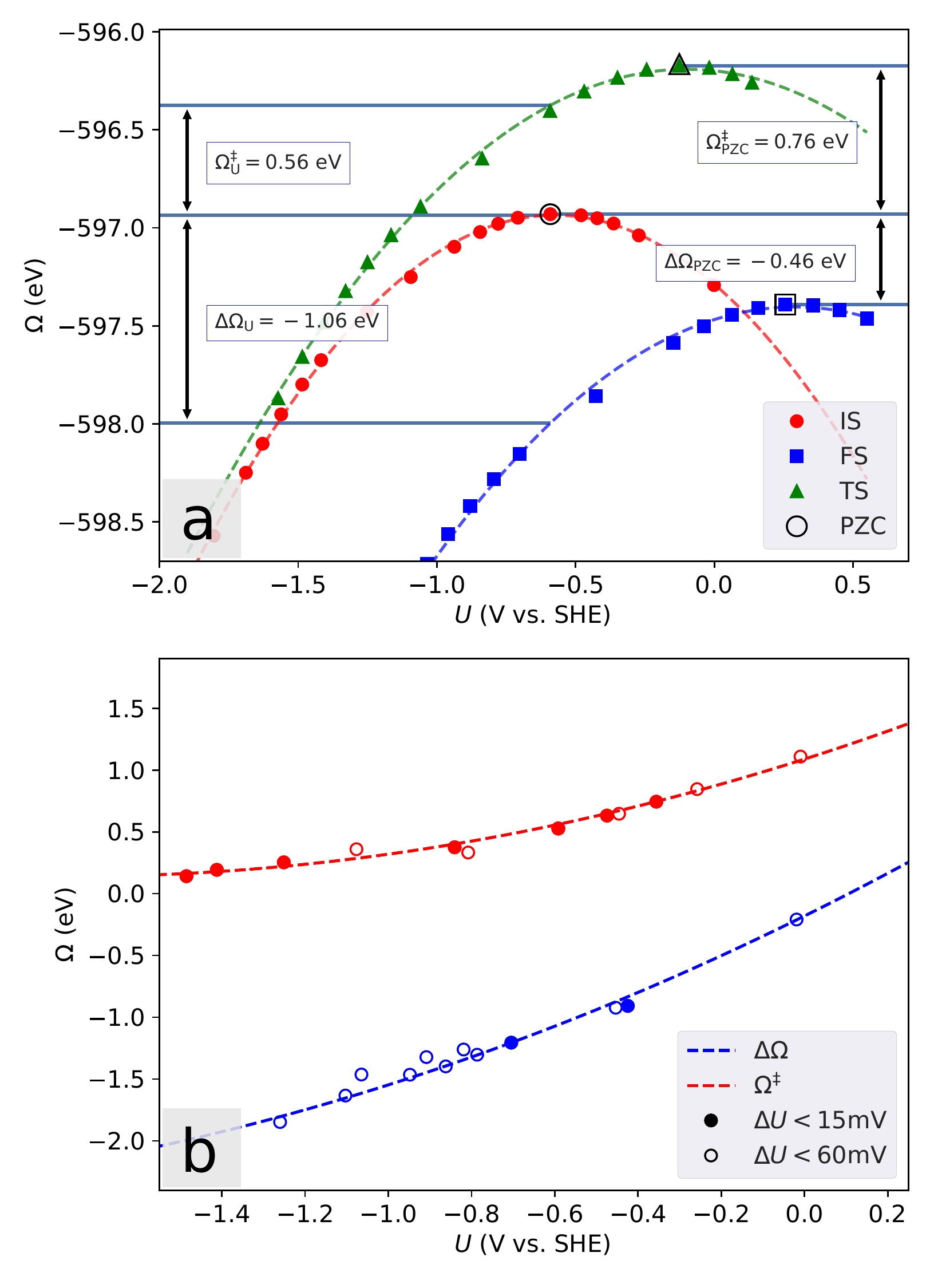}
    \caption{Example illustrating the methodology for converging reaction energy to a certain electrode potential, using parabolic fits. \textbf{a)} The neutral (PZC) energy is compared to that at fixed electrode potential $U=U_\mathrm{PZC}^\mathrm{IS}$. The neutral calculations do not correspond to a specific potential, as it changes over the reaction with a magnitude depending on the size of the simulated system. The reaction is seen to become increasingly exothermic at lower potentials, with corresponding lowering of the required activation energy. For large negative potential, the transition state (TS) converges towards the initial state (IS). \textbf{b)} Comparison of the obtained fits with the data points that are within a certain voltage threshold. The fit accurately reproduces the reaction and activation energy at a given potential. Data from the Heyrovsky reaction on \ce{Pt}-doped \ce{MoS2} with an initial hydrogen coverage of $\theta=2/3$.}
    \label{fig:volt_method}
\end{figure}

A finite simulation cell poses another challenge in determining the energetics of electrochemical reactions, as an electron transfer will lead to a substantial change in the electrostatic surface potential, and therefore to a capacitive contribution to the reaction energy. This contribution is inversely proportional to the lateral cell dimension, and the problem has previously been solved by extrapolating the energies to the limit of infinite lateral cell size \cite{ROSSMEISL200868}. This approach is quite expensive, and when dealing with constant dopant concentrations the cells rapidly become too large. Instead, we 
use a grand canonical approach (Figure \ref{fig:volt_method}) where the electron number is allowed to vary to adjust the voltage \cite{doi:10.1021/acs.jpcc.8b10046}.

Defining the electronic potential as $\mu = \epsilon_F + e\phi_\infty$, where $e$ is the 
electron charge, $\epsilon_F$ is the Fermi energy and $\phi_\infty$ is the electrostatic potential in the bulk solvent, the potential referred to that of the standard hydrogen electrode (SHE) is $U = \mu/e - \phi_\mathrm{SHE}$. For the SHE we use the value $\phi_{\mathrm{SHE}} = 4.43$~V \cite{1986417}. Further, we introduce the grand-canonical electronic energy $\Omega$ through the Legendre-transformation of the free energy $f$ as $\Omega_U = f(n_e) + \delta n_e e U$, where $\delta n_e$ is the number of excess electrons in the cell. 

Over a range in applied voltage that is not too large, the number of electrons is varied in increments starting from the neutral value, and the calculated energy of the system is then found to vary in a parabolic way with an extremum corresponding to the potential of zero charge (PZC). The curvature of the parabola represents the negative interfacial capacitance \cite{SANTOS200426}. As illustrated in Figure \ref{fig:volt_method}a, reaction and activation energies corresponding to a certain potential can then be obtained from the difference between fitted parabolas. The results shown in Figure \ref{fig:volt_method}b coincide well with the alternative (and more computationally expensive, unless only a specific potential value is of interest) approach of converging each reaction geometry to the same potential, though extrapolating the fits outside the potential range of the available data cannot be considered reliable. Most systems display excellent parabolic behavior, but in some cases changing the electron number leads to changes in the geometry or crossing of small electronic gaps. This causes an abrupt change in the Fermi energy, and correspondingly in the potential.

\section{Results}

\subsection{Doped \ce{MoS2}}
\begin{figure}[!h]
    \centering
    \includegraphics[width=\linewidth]{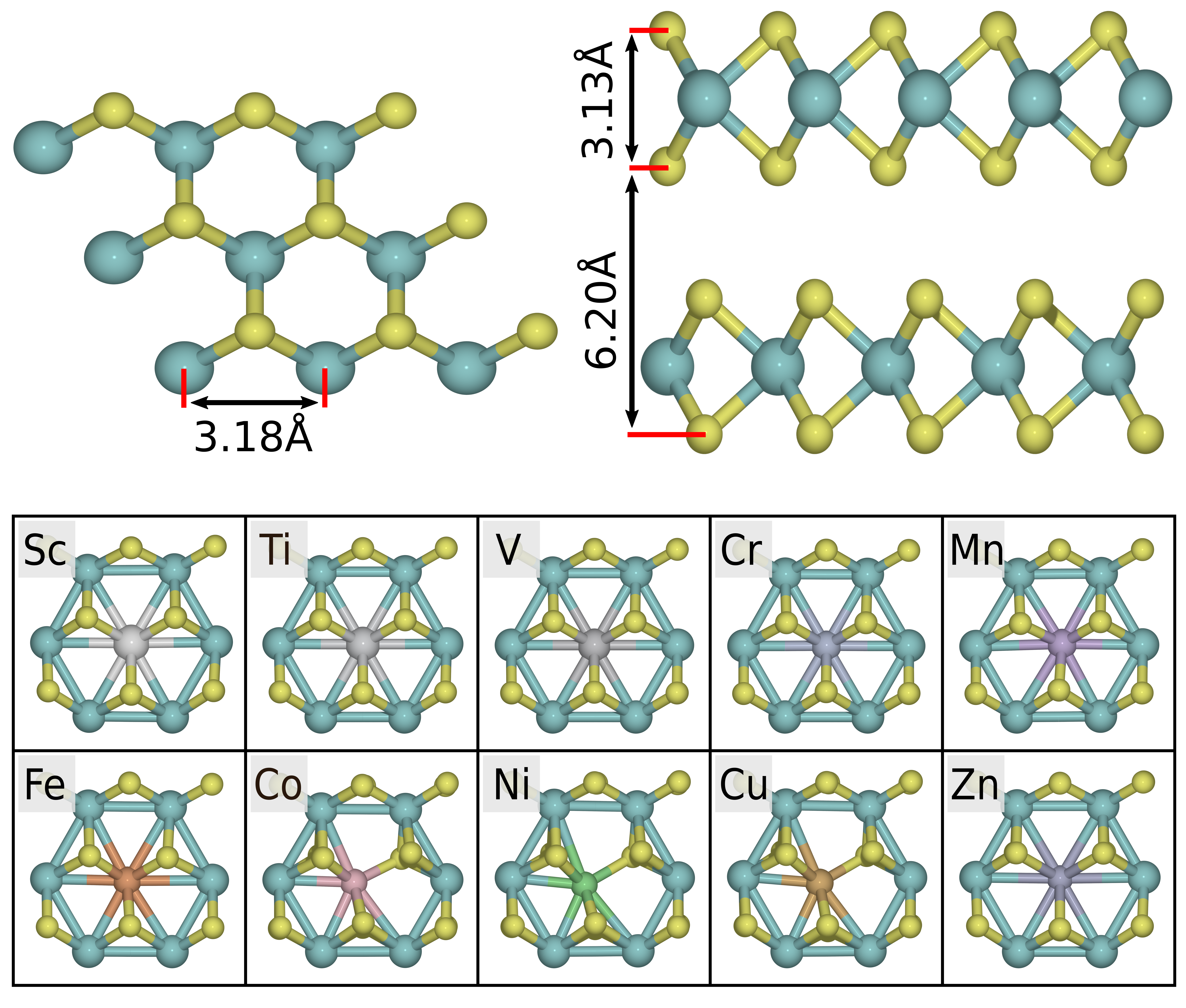}
    \caption{Top: Structure of pristine 2H-\ce{MoS2} with optimized lattice parameters. Bottom: Local geometries of \ce{MoS2} doped with 3d-transition metals. Most dopants retain the six-fold symmetry, but notably Co, Ni, and Cu break this symmetry.}
    \label{fig:geos}
\end{figure}

The bulk 2H-\ce{MoS2} phase consists of alternating layers, bound together by van der Waals forces, as shown in Figure \ref{fig:geos}. By minimizing the energy with respect to the unit cell volume, we obtain an optimized structure with in-plane lattice parameter $a=3.18$~{\AA}, interlayer separation $c=6.20$~{\AA}, and layer height $t=3.13$~{\AA}. Both the hexagonal lattice parameters are within $1\%$ of the experimental values ($a=3.15$~{\AA} and $c=6.15$~{\AA}, respectively \cite{PhysRevB.12.659}). Our calculations show that pristine 2H-\ce{MoS2} has band gaps of $\sim 0.9$~eV (indirect) and $\sim 1.6$~eV (direct) in multi- and mono-layer cases, respectively. This qualitatively corresponds to the experimental band structure, though the values are somewhat smaller due to the systematic underestimation of band gaps in the GGA approach. For comparison, the experimentally measured band gaps of bulk and monolayer 2H-\ce{MoS2} are roughly $1.3$~eV and $1.9$~eV \cite{PhysRevB.64.235305,PhysRevLett.105.136805}.

3d-transition metal atoms are introduced as \ce{Mo}-substitutional dopants to a $5\times5$ cell of the pristine monolayer, yielding a doping concentration of $4\%$. For \ce{Co}, \ce{Cu} and \ce{Ni}, the local symmetry of the pristine geometry is broken, and the dopant atom binds to five surrounding \ce{S}-atoms. For \ce{Sc}, \ce{Ti}, \ce{V}, \ce{Cr}, \ce{Mn}, \ce{Fe} and \ce{Zn}, the symmetry is preserved (see Figure \ref{fig:geos}). In the asymmetric case a sulfur atom has a broken or stretched bond, making it more exposed for adsorption (activated) with respect to the default coordination. We note that for \ce{Co}, \ce{Ni}, \ce{Cu}, \ce{Pt}, the dopant atom is also shifted down by $0.26-0.27$~{\AA} in the direction perpendicular to the plane (not visible in Figure \ref{fig:geos}). For \ce{Zn} there is a smaller shift of $0.15$~{\AA}. The remaining dopants are within $~0.01$~{\AA} of the reference height.

\begin{figure}[!h]
    \centering
    \includegraphics[width=0.9\linewidth]{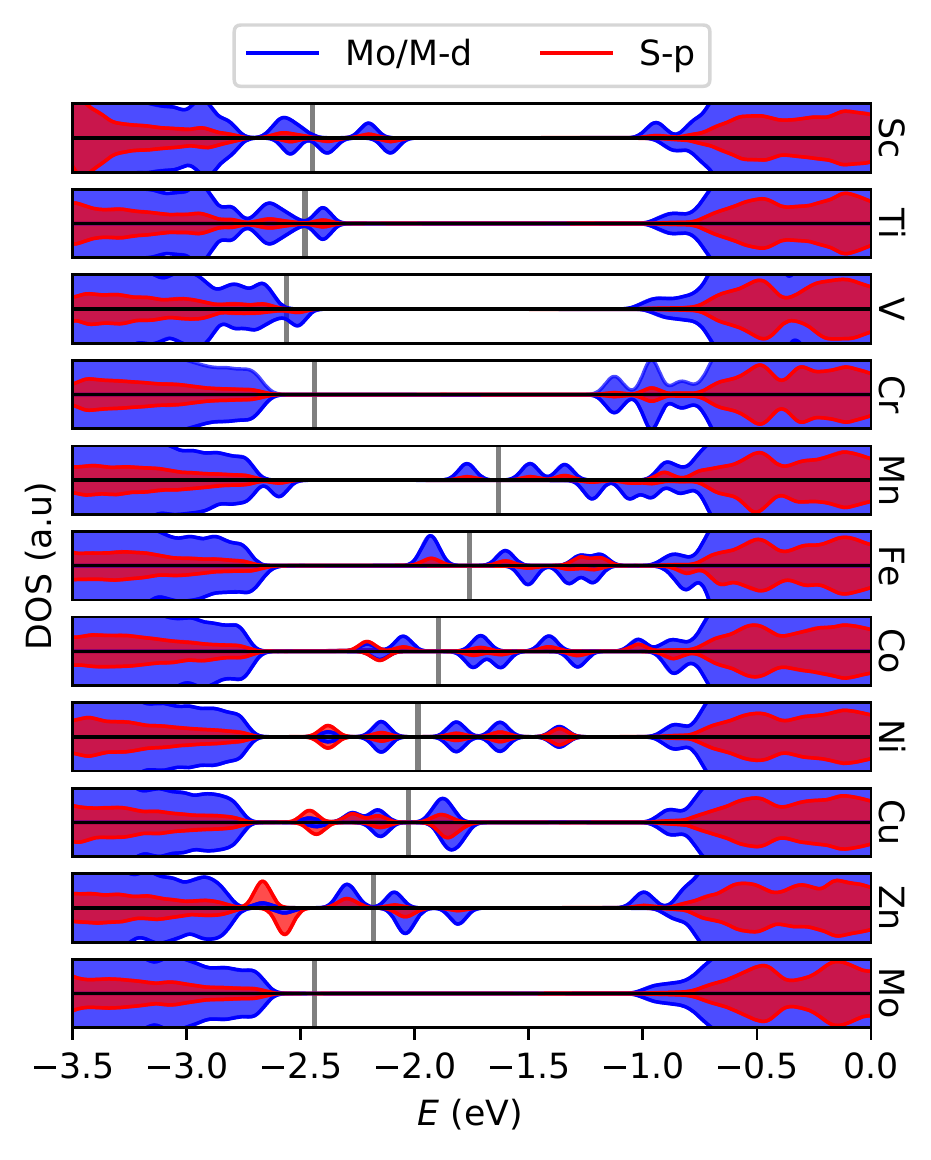}
    \caption{Projected density of states showing gap states emerging in 3d-metal doped \ce{MoS2} monolayers. Gray lines indicate the Fermi level.
    Pristine \ce{MoS2} in the bottom 
    graph
    for reference.}
    \label{fig:dos}
\end{figure}
 
 Electronic states within the band gap are introduced as a result of the doping.
 Figure \ref{fig:dos}
 shows
 the density of states (DOS) of the various systems near the band gap. Apart from \ce{Cr}, states are introduced in such a way that the effective band gap is significantly reduced. In the later 3d-transition metals, it appears that higher energy (occupied) \ce{S}-p states are introduced, consistent with activation. The induced states are mainly localized 
at 
the sulfur atoms neighboring the dopant atom, and in the asymmetric cases the dislocated atom contributes the most. DOS and its projections are detailed further for surrounding S in Figure S2$^\dag$. 

\subsection{Hydrogen Adsorption}
\begin{figure}[!h]
    \centering
    \includegraphics[width=0.9\linewidth]{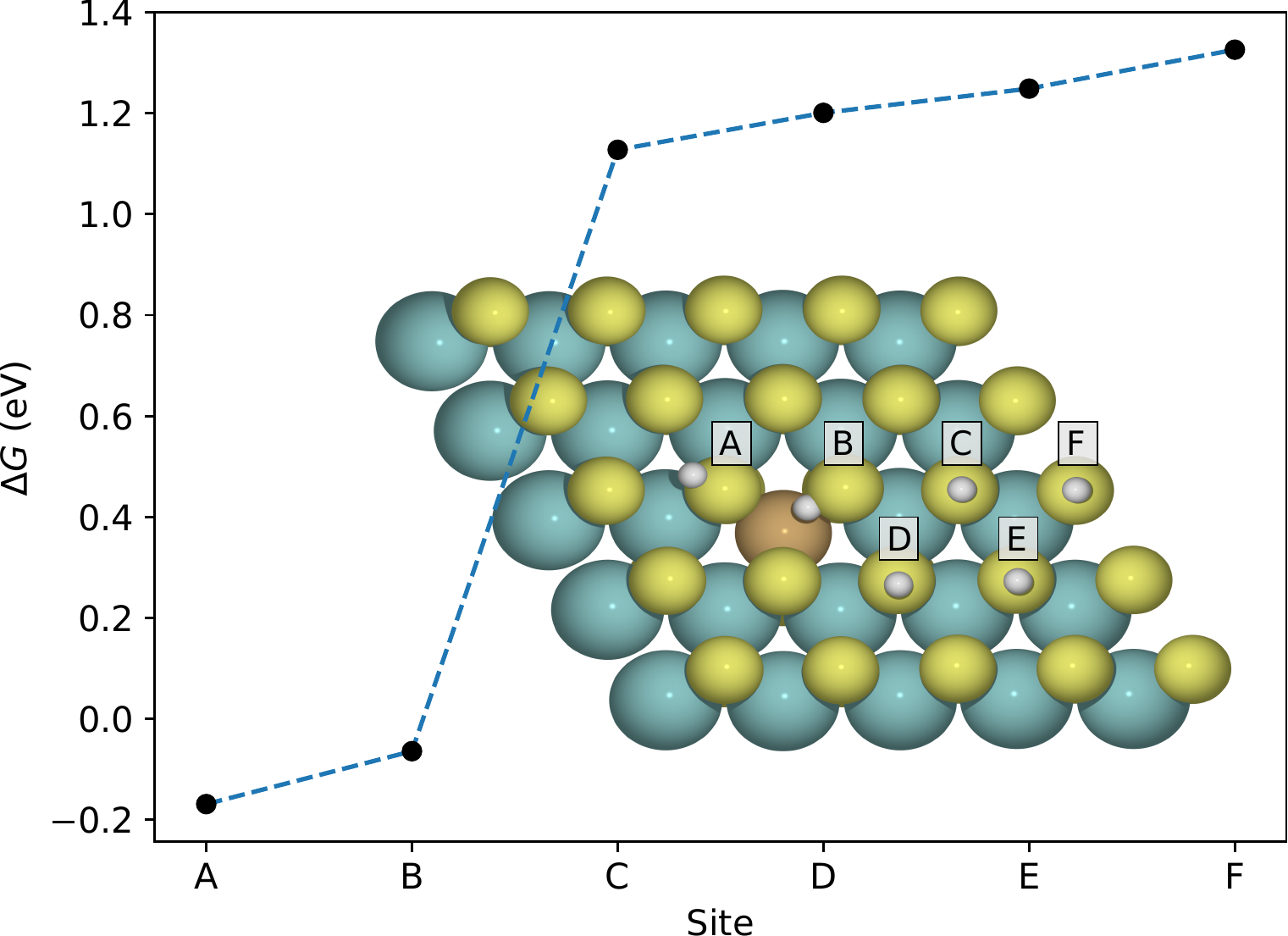}
    \caption{
Adsorption free energy of a single H atom coming from the gas phase and binding at
various sites on \ce{Cu}-doped \ce{MoS2}. The neighboring sulfur atoms are most notably activated. All sites display more favorable adsorption than the pristine basal plane.}
    \label{fig:ads-sites}
\end{figure}

As mentioned
above, the hydrogen adsorption free energy is a useful initial descriptor of the hydrogen evolution reaction, and therefore a natural starting point for investigation. We consider first the gas-phase situation, before moving on to solvated systems. Since hydrogen adsorption on the pristine basal plane is highly unfavorable, and due to the localization of the introduced \ce{S}-p states, we expect a low surface coverage where only the dopant-induced favorable sites near the dopant atom (and neighboring sulfur atoms) are occupied. Figure \ref{fig:ads-sites} shows the adsorption free energy for certain stable sites on \ce{Cu}-doped \ce{MoS2}. Only the sites in immediate vicinity of the \ce{Cu}-atom (A and B) are favorable, and we will focus mainly on these sites in the following. This trend is similar in the remaining 3d metals, and the neighboring sites are lowest in energy also in the cases where adsorption is unfavorable. For details we refer to Figure S3$^\dag$.
 
\noindent The free energy is obtained from the electronic energy as explained in the previous section. The ZPE correction 
is
calculated for a 
single
\ce{H}-adsorption for each system. Calculated $E_\mathrm{ZPE}$ for the dopant systems span a range of only $0.015$~eV, and 
it can be approximated 
as a constant correction.
In a similar manner, we assume that the vibrational frequencies do not change considerably 
as the coverage increases.

\begin{figure}[!h]
    \centering
    \includegraphics[width=\linewidth]{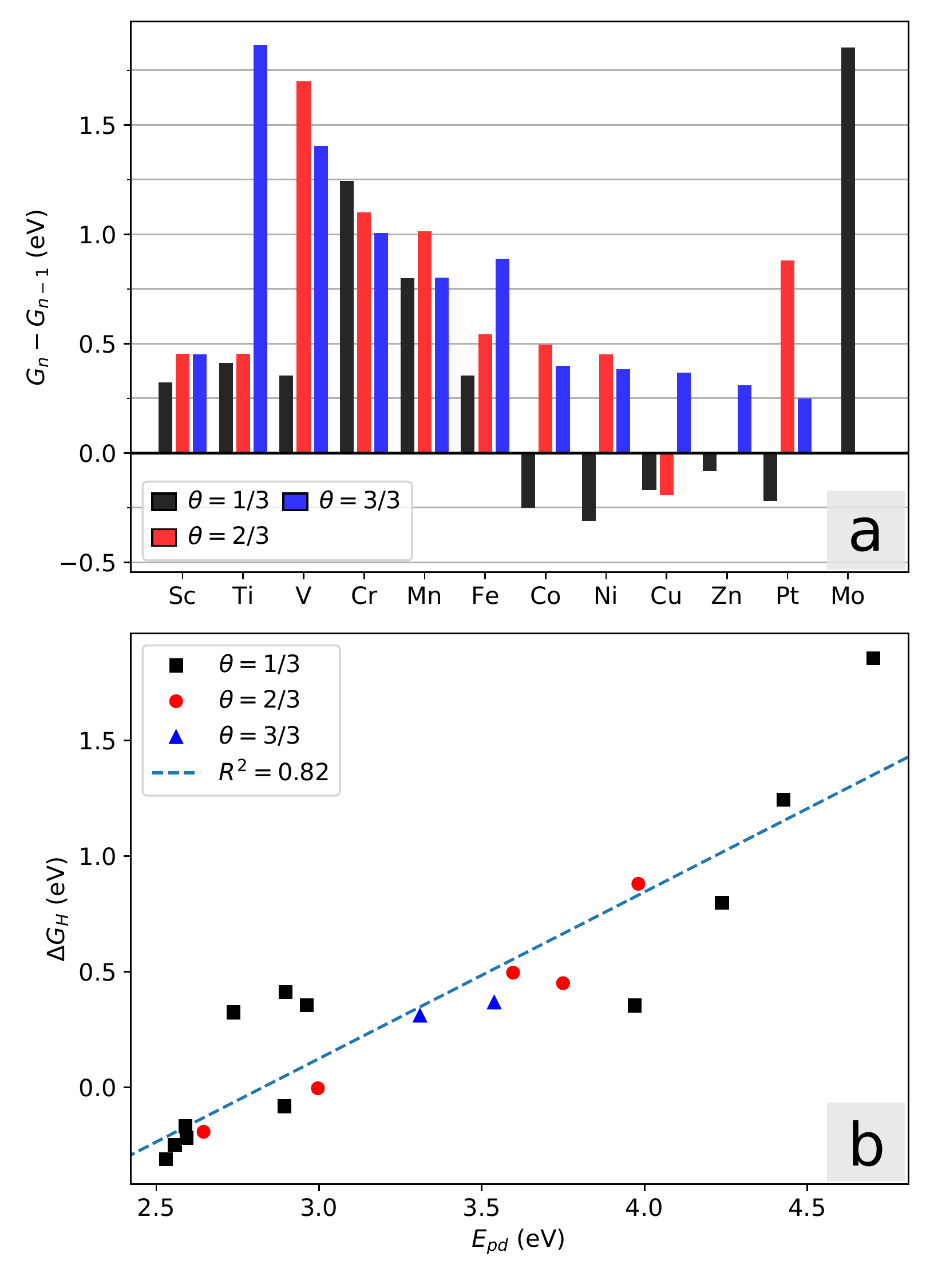}
    \caption{\textbf{a)} Incremental free energy of adsorption of multiple \ce{H}-atoms for the doped and pristine \ce{MoS2}. \textbf{b)} The promotion energy $E_\text{pd}$ vs. free energy of adsorption for coverages up to the first unfavorable adsorption for each dopant.}
    \label{fig:ads}
\end{figure}

The free energy of adsorption for one to three \ce{H} 
atoms in the simulation cell
is given in Figure \ref{fig:ads}a. Consistent with the previous arguments of symmetry and electronic DOS, we observe that only the \ce{Co}, \ce{Ni}, \ce{Cu} and \ce{Zn} systems show energetically favorable adsorption of a single hydrogen atom, 
and
higher coverages are even less favorable. Adsorption configurations that display moderate change in free energy are of interest for hydrogen evolution. This means that for systems \ce{Sc} through \ce{Fe}, only single H adsorption is relevant (note that \ce{Cr} and \ce{Mn}
have high adsorption 
energies in any case).
For systems \ce{Co} through \ce{Zn}, we consider an initial coverage of one or two \ce{H} per dopant atom. For reference, the first adsorption energy was also calculated for the two-layer MoS$_2$ case of systems \ce{Fe} through \ce{Zn}, with only the top layer doped. Compared to the monolayer case, these values differ by less than $0.02$ eV, suggesting that the hydrogen binding onto these dopant-activated sites is not sensitive to the presence of underlying layers. Monolayer calculations are thus appropriate for representing the general slab systems. \\

\noindent A chemical picture of the adsorption mechanism on \ce{MoS2} is that excess charge on the sulfur atom upon adsorption is partially distributed over the surrounding metal atoms \cite{https://doi.org/10.1002/anie.202003091}. In this regard, the promotion energy is a useful descriptor for the adsorption energy: $E_\mathrm{pd} = E_\mathrm{d}-E_\mathrm{p}$, where $E_\mathrm{p}$ is the p-orbital center of the \ce{S}-atom in question, and $E_\mathrm{d}$ is the center of the unoccupied metal d-orbitals integrated from the Fermi level up to the point where one electron is added, i.e. the effective LUMO of the local system. Figure \ref{fig:ads}b shows the correlation between the adsorption energy and $E_\mathrm{pd}$. We note that the local geometry changes upon hydrogen adsorption, which leads to additional energy contributions and deviations from this descriptor. This is especially visible for the \ce{Fe}-system, where the local six-fold symmetry is broken upon hydrogen adsorption. The other systems maintain their symmetry, but the distance between the dopant atom and the active sulfur changes. The promotion energy captures two important conditions that are necessary for favorable adsorption energy, namely that the sulfur p-orbital states must be high in energy (activated), and that the surrounding d-orbitals must have unoccupied states not far above the Fermi level. It follows that large band gaps are detrimental towards favorable adsorption in these systems.

\subsection{Neutral Cell Hydrogen Evolution}
\begin{figure}[!h]
    \centering
    \includegraphics[width=\linewidth]{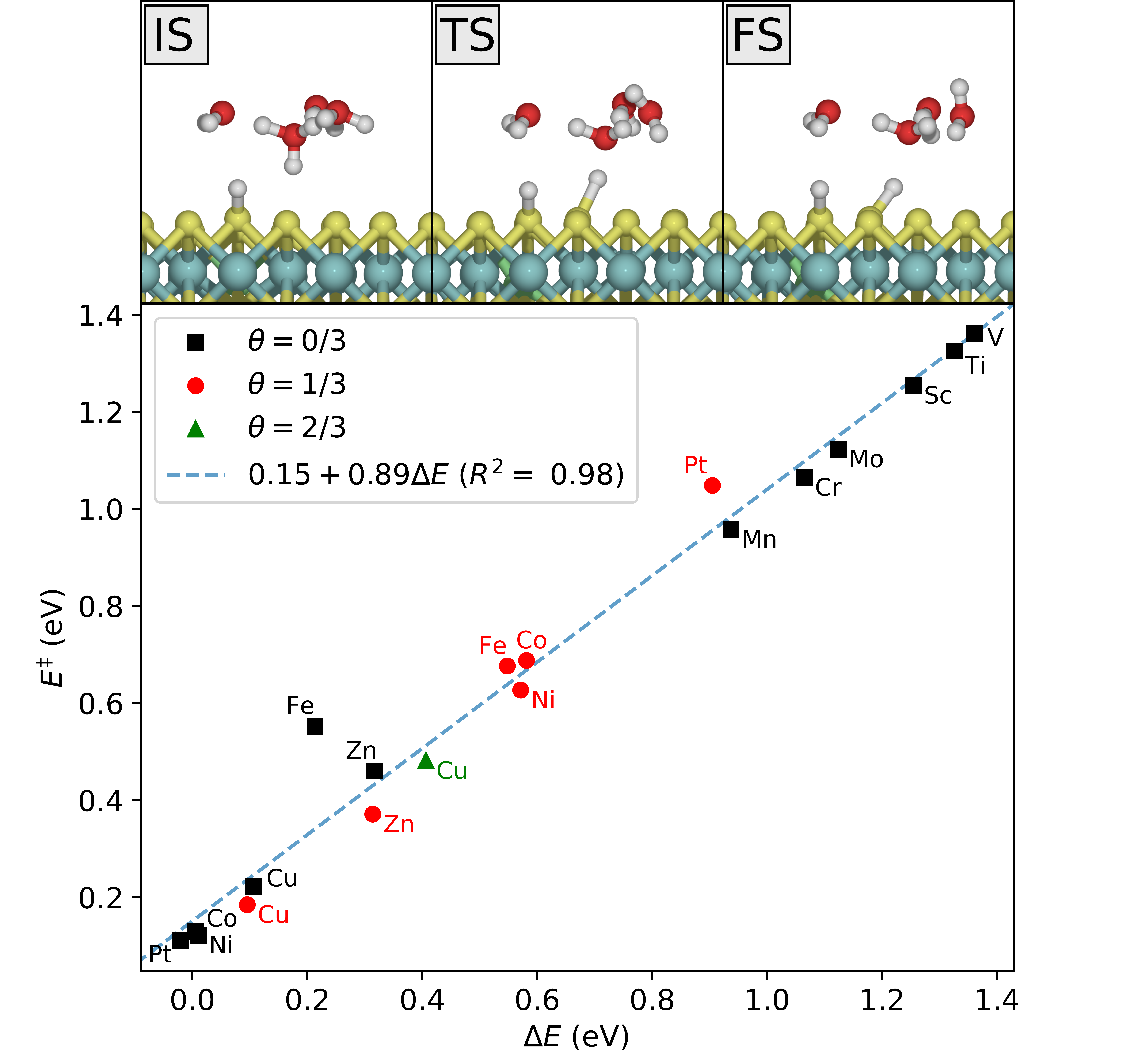}
    \caption{Top: Reaction mechanism for the Volmer reaction step on \ce{Ni}-doped \ce{MoS2} at an intial coverage of $\theta = 1/3$. 
    Bottom: 
    Calculated activation energy vs. reaction energy for the various
    dopants and coverages studied here at the potential of zero charge, and comparison with the 
    Brønsted-Evans-Polanyi relation (dashed line).
    }
    \label{fig:bep_volm}
\end{figure}

Next, 
the presence of the solvent is taken into account
and 
the various
steps of the hydrogen evolution 
are calculated
where the systems are kept under neutral conditions, that is at the potential of zero charge (PZC). Afterwards the grand-canonical approach will be used to keep the electrode potential fixed during the reaction. We consider first the Volmer step, where a proton from the water cluster
adsorbs
on the \ce{MoS2} surface. It follows a simple reaction path, depicted in Figure \ref{fig:bep_volm} for the \ce{Ni}-doped system at initial coverage $\theta = 1/3$. As also demonstrated in Figure \ref{fig:bep_volm}, the relation between activation and reaction energies for the Volmer reaction on the doped systems follows the Brønsted-Evans-Polanyi principle, where the activation energy $E^\ddag$ is linearly proportional to the reaction energy. The Volmer step is highly unfavorable 
for
all the early 3d-metals, even though \ce{Sc}, \ce{Ti}, and \ce{V} have much lower adsorption energies in the gas phase. For \ce{Sc}, \ce{Ti}, \ce{V}, \ce{Cr} and \ce{Mo}, there appears to be either no saddle point (transition state, TS) between the initial (IS) and final state (FS), or the TS is very close to the FS, as obtained with both MMF and CI-NEB searches. Thus, the activation energy tends towards the reaction energy as the reaction becomes more endothermic, and the Volmer reaction is effectively only uphill in energy in these systems. Importantly, the adsorption state is not kept (meta)stable by a reverse barrier and its lifetime is insignificant.

The large difference between the gas phase adsorption energy and the Volmer reaction energy in solution obtained for some of the systems (in particular \ce{Sc}, \ce{Ti}, \ce{V}) shows that the solvent plays an important role in the reaction energetics. That is, the water cluster interacts differently with the surface before and after its proton has been transferred.
For example, for systems with favorable hydrogen adsorption, the cluster is weakly bonded to the adsorbed hydrogen through a hydrogen bond. Such a configuration is not stable for the systems with unfavorable hydrogen adsorption, as it would lead to the hydrogen atom re-entering the solution. The cluster is thus further from the surface in a metastable final state, not supporting the attractive interaction. The correspondence with gas phase adsorption becomes clear again if one considers the next step of the reaction. After the Volmer step, the water cluster is again supplied with a proton from the bulk acidic solvent. The difference in energy between this state and the initial Volmer state (correcting for the additional \ce{H} atom), is what corresponds to the adsorption energy. However, what influences the Volmer barrier is the initial Volmer reaction energy, as is clear from Figure \ref{fig:bep_volm} and the linear relation. This illustrates one aspect of the significance of the choice of solvent description. We 
note that there is a small energy barrier of roughly $0.15 \,\mathrm{eV}$ associated with the proton transfer step from the bulk solution. This is an upper bound in the sense that proton diffusion towards a negative surface will be associated with a favorable free energy change given by the potential difference vs. the bulk solvent, thus also lowering the associated barrier.

\begin{figure}[!h]
    \centering
    \includegraphics[width=\linewidth]{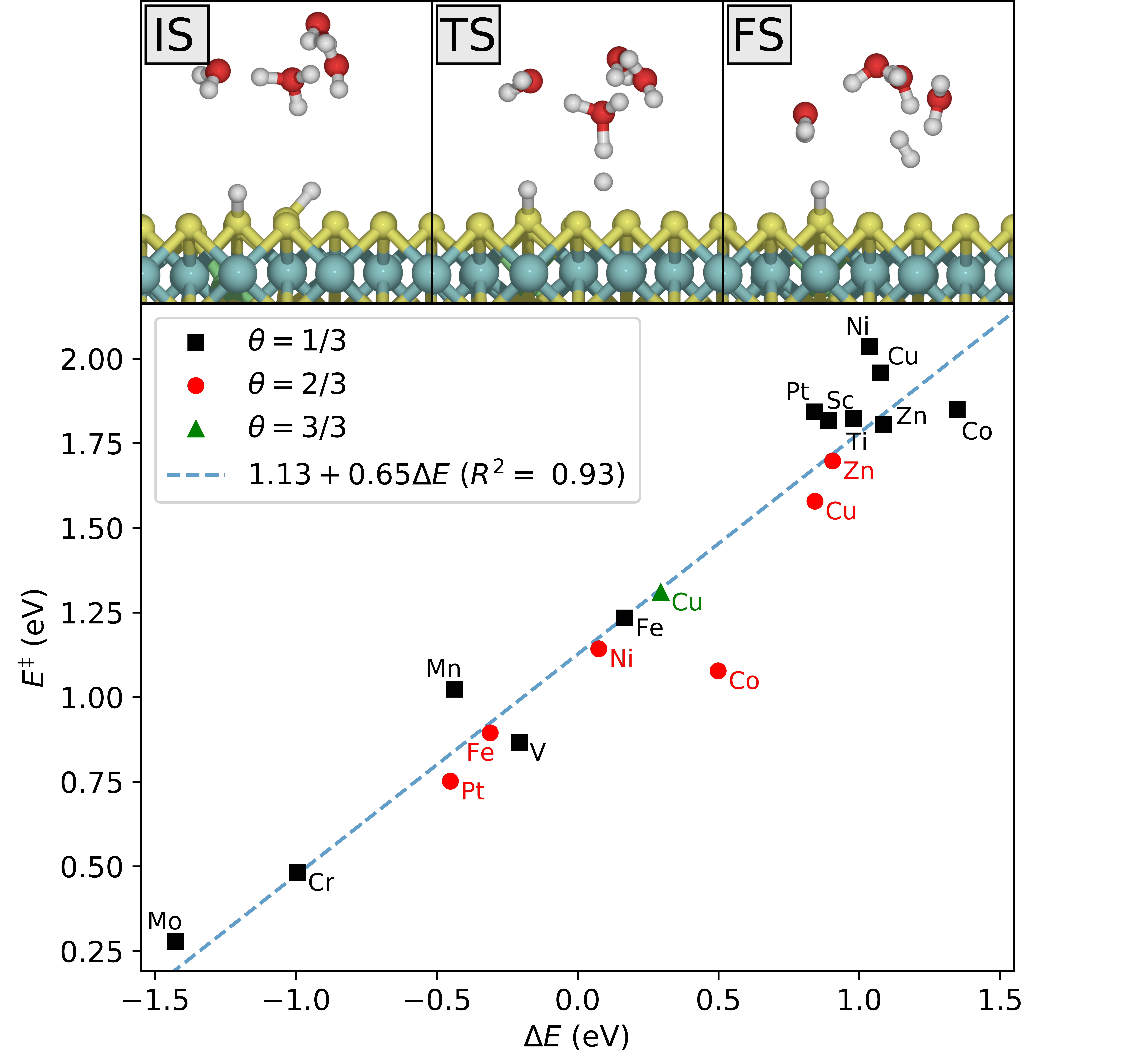}
    \caption{Top: Reaction mechanism for the Heyrovsky reaction step on \ce{Ni}-doped \ce{MoS2} at an intial coverage of $\theta = 2/3$. 
    Bottom: 
    Calculated activation energy vs. reaction energy for the various
    dopants and coverages studied here at the potential of zero charge, and comparison with the 
    Brønsted-Evans-Polanyi relation (dashed line).
    }
    \label{fig:bep_heyr}
\end{figure}

In the Heyrovsky step, a proton in the water cluster binds to an adsorbed hydrogen atom, and 
an electron is transferred from the electrode to form an \ce{H2} molecule that is released
from the surface.
The reaction path is shown in Figure \ref{fig:bep_heyr}. As in the Volmer reaction, the activation energy follows a linear trend with respect to the reaction energy. In this case, the four last 3d-metals, as well as \ce{Sc} and \ce{Ti}, are seen to have highly unfavorable Heyrovsky reaction energies at low coverage. For the late 3d-metals this is expected from the Volmer energies, but \ce{Sc} and \ce{Ti} appear to have highly unfavorable energies for both reactions. This is related to the previous discussion on the distinction between the initial Volmer reaction and the following proton resupplying step. Once resupplied, \ce{Sc} and \ce{Ti} return to lower energies, while \ce{V} does not. This is consistent with the obtained Heyrovsky energetics. 

The Tafel step involves the desorption of two adsorbed \ce{H}-atoms on the surface. The kinetics of this step therefore 
depend heavily
on the geometries and relative energies of the adsorption sites, as well as the surface \ce{H}-coverage. 
This step depends less on the presence of the solvent 
than the Volmer and Heyrovsky steps. 
Adsorption of several H atoms
will naturally involve occupying the less favorable sites, as well as introducing \ce{H}-\ce{H} interactions. From the single-\ce{H} adsorption energies, we expect that subsequent adsorption will still be confined around the dopant atom, although energetically less favorable due to \ce{H}-\ce{H} repulsion. As shown in Figure S6$^\dag$, it is evident that the dopant-induced sites are still preferred despite 
repulsive 
\ce{H}-\ce{H} interaction. 
Referring to the labeling of the sites in Figure \ref{fig:ads-sites}, the
optimal reaction path proceeds from a neighboring A-A configuration through the A-B and B-B intermediates.
For details on the reaction path and resulting activation energies and scaling relation, we refer to Figures S7 and S8$^\dag$. 

Comparison of the activation energies of the Tafel step and the Heyrovsky step at the same hydrogen coverage ($\theta = 2/3$) shows that HER is more likely to occur by a Volmer-Heyrovsky mechanism rather than a Volmer-Tafel mechanism, for all dopants except \ce{Zn}. In these systems the barrier of the Tafel mechanism is significantly larger than for the Heyrovsky mechanism at the same initial coverage, but for \ce{Zn}, the Heyrovsky and Tafel barriers are comparable, at $1.70$ and $1.67$~eV, respectively.

\subsection{Constant Electrode Potential}
So far, neutral systems at the PZC have been considered. We investigate next the energetics when the reaction occurs at a certain constant potential by varying the number of electrons 
as explained earlier. Fixing the electrode potential will lead to a certain correction to the PZC energies as demonstrated in Figure \ref{fig:volt_method}a. The preferred reaction mechanism is not expected to change for most systems, but in general it will be a function of the applied potential. The Volmer-Heyrovsky process is expected to become more favorable the more negative the applied potential is, while the Tafel step is only weakly affected. Calculations for \ce{Fe} through \ce{Zn} confirm that the Tafel barrier potential dependence in general is weaker than that of the electron-transfer processes. 
For \ce{Zn}$_{\theta=2/3}$, the Tafel energy barrier is lower than the Heyrovsky barrier above $-0.8$~V, but such a high equilibrium coverage, $\theta=2/3$, is only reached at voltage below $-0.5$~V, so the Volmer-Tafel path will thus be (slightly) preferred in a range within this voltage window, but otherwise the Volmer-Heyrovsky path is preferred. For all other systems, the Volmer-Heyrovsky path remains more favorable at all potentials. With this in mind, we consider for simplicity the Volmer-Heyrovsky path of all systems in the following comparison.

Around $0$~V vs. SHE, the adsorption energies are quite similar to those in the gas phase, and the configurations we noted earlier are still of interest. Therefore, we consider the initial hydrogen coverages of 0 (\ce{Sc}, \ce{Ti}, \ce{V}, \ce{Cr}, \ce{Mn}, \ce{Fe}, \ce{Mo}), 1 (\ce{Co}, \ce{Ni}, \ce{Zn}, \ce{Pt}), and 2 (\ce{Cu}) per dopant atom in equilibrium. For reference, the initial zero-coverage is also considered for all dopants.

\begin{figure}[!h]
    \centering
    \includegraphics[width=\linewidth]{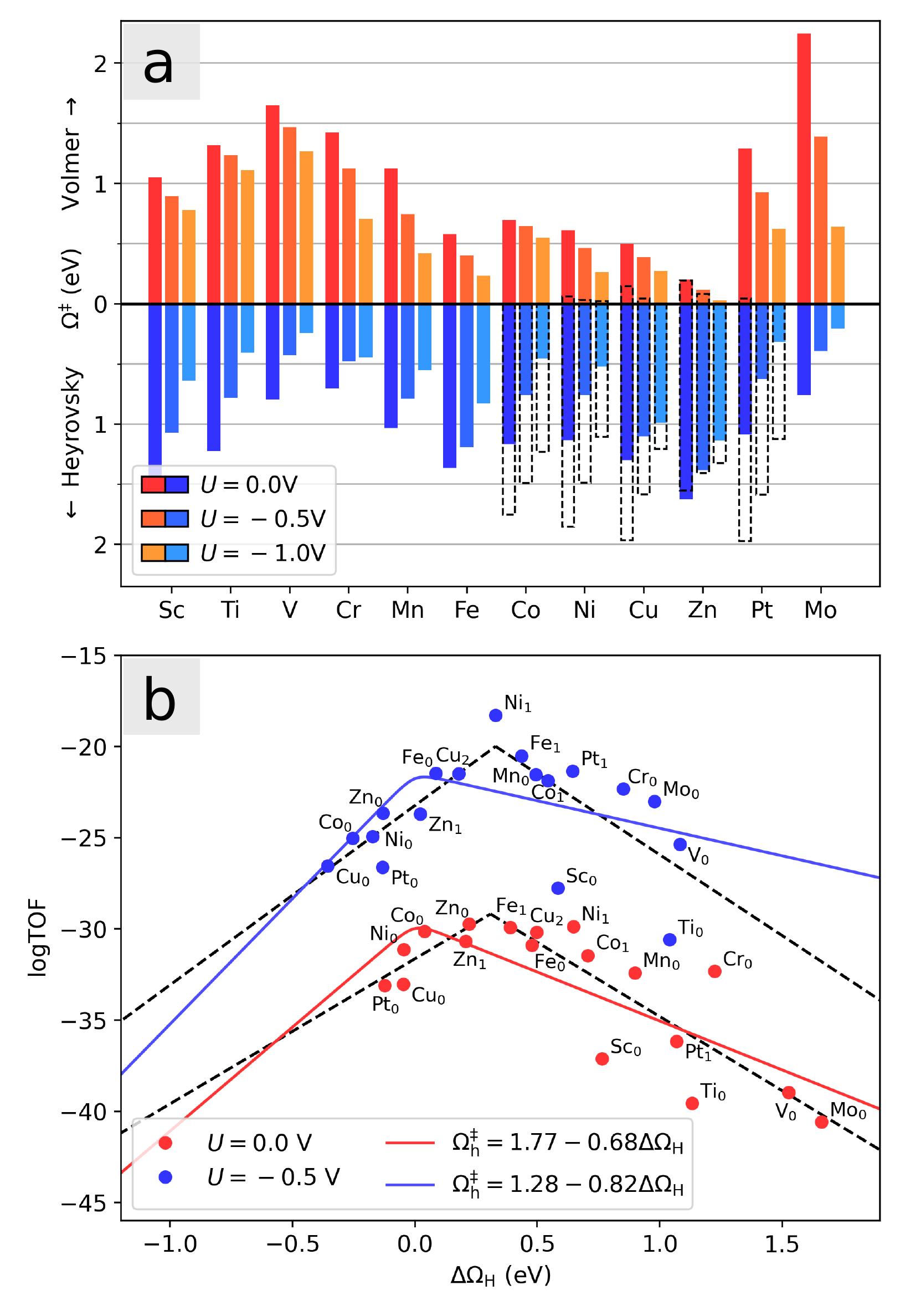}
    \caption{\textbf{a)} Activation barriers for the Volmer and Heyrovsky steps at $U=0$~V, $U=-0.5$~V and $U=-1.0$~V for the various dopants. 
    The initial H-adatom coverage is taken to be the
    equilibrium coverage at $U=0$~V. 
    Dashed bars indicate the values for the case of $\theta = 0$ for systems \ce{Co}, \ce{Ni}, \ce{Cu}, \ce{Zn} and \ce{Pt}. Since the first adsorption is favorable in those systems, it is seen to be associated with a 
    small Volmer barrier, and a correspondingly large Heyrovsky barrier. 
    \textbf{b)} Relation between the activity obtained from a simple kinetic model and $\Delta\Omega_\mathrm{H}$ for each studied dopant and coverage at $U=0$~V and $-0.5$~V. Dashed lines show fits of the calculated points, while solid lines show the theoretical activity calculated from the relation between the Heyrovsky barrier $\Omega_h^\ddag$ and hydrogen adsorption energy $\Delta\Omega_\mathrm{H}$. The slope of this relation determines the resulting volcano shape.}
    \label{fig:barrs}
\end{figure}

Figure \ref{fig:barrs}a shows the potential dependence of the Volmer and Heyrovsky barriers for each investigated system in the range between $0$~V and $-1$~V, evaluated at the equilibrium coverage at $0$~V. 
Both barriers are lowered by the negative applied potential. Most notably, we observe that the \ce{Co}- and \ce{Ni}-doped systems with an initial hydrogen atom display significantly smaller Heyrovsky barriers than the rest of the systems with small Volmer barriers, especially around $U=-0.5$~V. Since balanced moderate barriers in general will lead to faster reaction kinetics than one small and one large barrier, these systems seem to have the most active basal planes for HER. 
Note again that to a first approximation, the adsorption energy is modified by $eU$ (see Equation \ref{mu_H}) such that higher coverages will be favorable at large negative applied potentials. For $U=0$~V and $U=-0.5$~V, endothermic adsorptions are included for all systems as seen in Figure \ref{fig:barrs}b, ensuring that the relevant coverages are considered. In the case of $U=-1$~V, the large potential could lead to a further increase of the H-coverage, which here would lead to qualitative increase (decrease) in the calculated Volmer (Heyrovsky) barrier. The pristine \ce{MoS2} 
shows a particularly strong dependence on the applied voltage
and surprisingly exhibits low barriers at large negative potential ($U=-1$~V). 

A detailed evaluation of reaction kinetics would require
calculations of several other parameters, such as pre-exponential factors. 
However, using the fact that all the systems studied are comparable, we can 
estimate 
the relative kinetics with a simple kinetic model. We characterize the turnover frequency (TOF) $f$ by $f_i \propto p_i e^{-\Omega_h^\ddag/k_\mathrm{B}T}$ where $p_i = \mathcal{Z}^{-1}{e^{-\Delta \Omega_\mathrm{H_i}/k_\mathrm{B}T}}$ is the probability of being in the state with $i$ adsorbed hydrogen and $\Omega^\ddag_h$ is the Heyrovsky barrier. $\mathcal{Z}$ is the partition function. At temperature $T=298$~K, the resulting TOF assumes the characteristic volcanic shape with respect to $\Delta \Omega_\mathrm{H}$, as seen in Figure \ref{fig:barrs}b. Points are calculated directly from the 
energetics for
each system, while the solid lines represent the theoretical activity obtained by using the scaling relation between $\Delta \Omega_\mathrm{H}$ and $\Omega^\ddag_h$ at the corresponding potential. This correlation is shown in Figure S9$^\dag$, where the same outliers as in Figure \ref{fig:barrs}b are visible, notably  \ce{Sc}, \ce{Ti}, \ce{Cr}, \ce{Ni} at $U=0$~V and \ce{Sc}, \ce{Ti}, \ce{Ni} at $U=-0.5$~V. These outliers lead to the different shapes between the two volcanoes. Peaks of the dashed volcanoes are shifted towards positive $\Delta\Omega_\mathrm{H}$, indicating that in these systems the optimal condition is a slightly unfavorable adsorption rather than the perfectly neutral one, due to the competition between the Volmer and Heyrovsky barriers not being balanced at $\Delta\Omega_\mathrm{H} = 0$~eV.

\begin{figure}[!h]
    \centering
    \includegraphics[width=\linewidth]{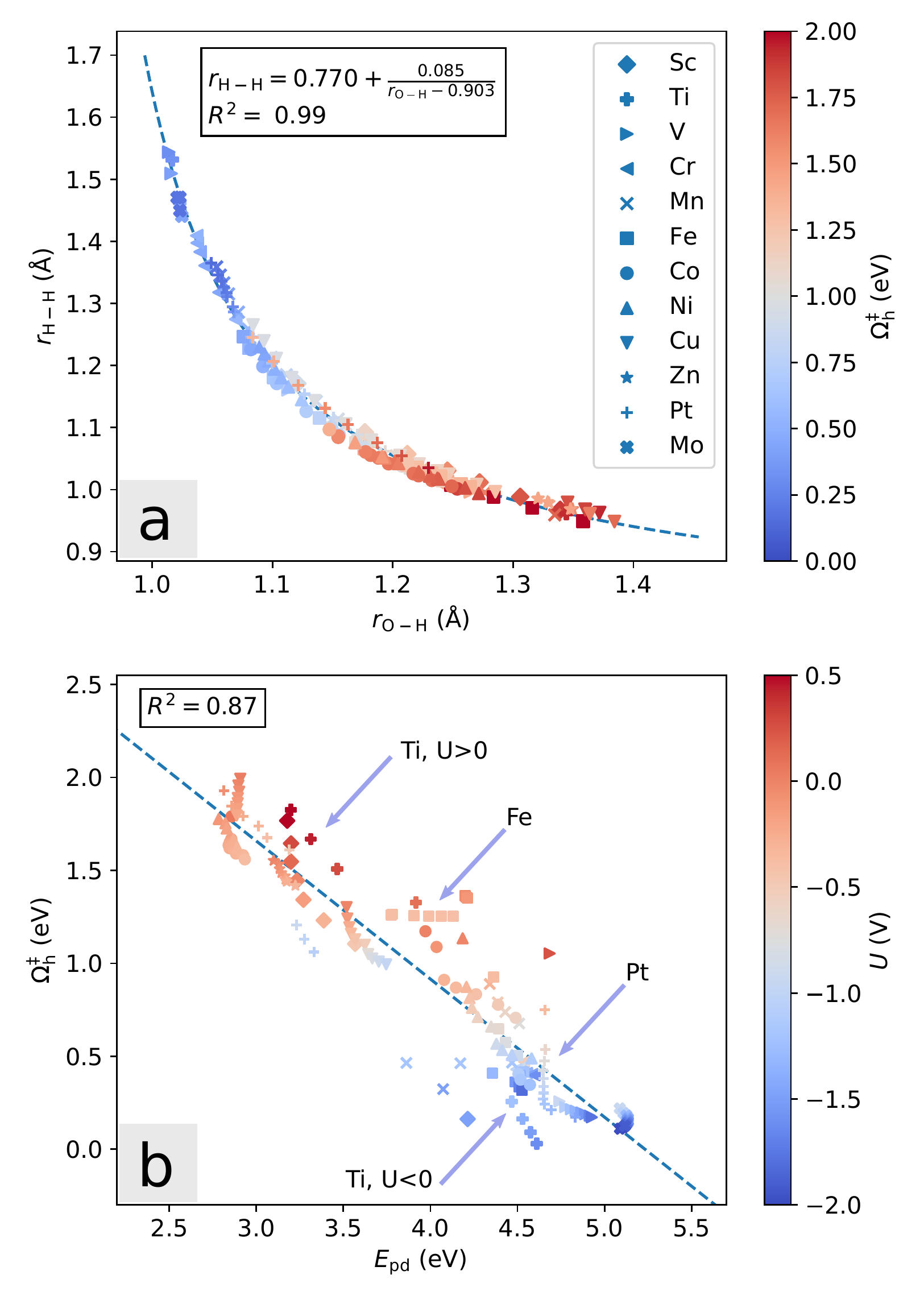}
    \caption{\textbf{a)} Correlation of the distances $r_{\mathrm{O}-\mathrm{H}}$ and $r_{\mathrm{H}-\mathrm{H}}$ at the transition state, and the Heyrovsky activation energy $\Omega_\mathrm{h}^\ddag$ for all dopants at various values of potential and hydrogen coverage. \textbf{b)} Correlation of the promotion energy $E_\mathrm{pd}$ before hydrogen adsorption and $\Omega_\mathrm{h}^\ddag$ at the corresponding potential and hydrogen coverage. The promotion energy captures the main trend of $\Omega^\ddag$. Some particular features are exemplified by the \ce{Fe}-system (horizontal), \ce{Pt}-system (vertical), and \ce{Ti}-system (gap crossing). }
    \label{fig:r_corr}
\end{figure}

Further, we note that the transition state geometries change slightly as a function of the potential. For the Heyrovsky reaction, the distances $r_{\mathrm{H}-\mathrm{H}}$ (between the two reacting hydrogen atoms), $r_{\mathrm{O}-\mathrm{H}}$ (between oxygen atom and proton), and $r_{\mathrm{S}-\mathrm{H}}$ (between sulfur atom and adsorbed hydrogen atom) largely define the reaction coordinate. As the potential is lowered, the magnitude of these values at the transition state tend to increase, decrease and decrease, respectively. In Figure \ref{fig:r_corr}a, this is illustrated for all systems considered. Regardless of the dopant, a low Heyrovsky barrier is associated with a transition state which is geometrically more similar to the initial state than the final state, in that $r_{\mathrm{O}-\mathrm{H}}$ is close to the \ce{H3O+} bond length and $r_{\mathrm{H}-\mathrm{H}}$ is far from the \ce{H2} bond length. Thus, knowledge of the 
arrangement of the water molecules
at the transition state is largely indicative of the activation barrier. The analytical fit has asymptotes at $r_{\mathrm{H}-\mathrm{H}} = 0.770$~{\AA} and $r_{\mathrm{O}-\mathrm{H}} = 0.903$~{\AA}. 

In terms of electronic structure, the previously discussed promotion energy $E_\mathrm{pd}$ captures the overall picture well. $E_\mathrm{pd}$ is positively correlated with adsorption energy, and relates to the Heyrovsky barrier as seen in Figure \ref{fig:r_corr}b. Interestingly, the relation within each individual system does not necessarily follow the overall trend. This illustrates the different contributions to the activation mechanism (chemical adsorption energy as described by $E_\text{pd}$, and the electrode potential), and is especially visible for the \ce{Mn} and \ce{Fe} systems. These systems are hexagonally symmetric at PZC, but the ground state symmetry is spontaneously broken once a certain amount of excess negative charge is introduced. Additional charge further increases the distance between the dopant atom and the activated \ce{S}, but during this transition the Fermi level (and hence potential) remains nearly constant, as the energies of unoccupied \ce{S}-$p$ states are lowered simultaneously with introduction of more electrons. In Figure \ref{fig:r_corr}b this manifests as a local horizontal trend. For \ce{Fe} and \ce{Mn}, the changes in $r_\mathrm{M-S}$ are roughly $0.20$~{\AA} and $0.25$~{\AA}, respectively. Comparing with vertical trend systems such as \ce{Cu} and \ce{Pt}$_{\theta=1/3}$, we find changes of roughly $0.015$~{\AA}. Within each system, $E_\mathrm{pd}$ is thus closely related to the distance $r_\mathrm{M-S}$ and does not seem to otherwise depend strongly on the amount of excess charge, unless it leads to the crossing of small gaps. Large change in $E_\mathrm{pd}$ due to gap crossing without significant geometrical change can be seen in Figure \ref{fig:r_corr}b for the \ce{Sc} and \ce{Ti} dopants. In systems with small change in the promotion energy, the potential is the main contribution to barrier decrease. 

\section{Discussion}
In all cases, single-atom doping results in a lowering of the PZC hydrogen adsorption energy. For the later 3d-metals, the doping results in breaking the local hexagonal symmetry, which leads to relatively larger affinity of the sulfur atom to hydrogen adsorption. This further leads to a wide range of HER-behavior for the dopants, from Volmer-limited to Heyrovsky-limited and even Tafel-limited in the case of \ce{Zn} at moderate applied potential. Within the setting of the basal plane, particularly \ce{Co} and \ce{Ni} stand out with a good balance of the two barriers at moderate applied potential, though according to the kinetic model \ce{Ni} is clearly more active at $-0.5$~V. At that point, the \ce{Ni}-doped system exhibits barriers of $0.47$ and $0.76$~eV. In comparison, the Volmer and Tafel barriers on \ce{Pt}(111) have with similar methodology been found to be $0.66$ and $0.55$~eV, respectively, at $0$~V and a full monolayer \ce{H}-coverage \cite{doi:10.1021/acs.jpcc.8b10046}. Therefore, the performance of the doped basal plane will be (at best) comparable to monocrystalline \ce{Pt}(111) at a $0.5$~V 
more negative
applied potential. However, experiments often display comparable exchange currents at significantly lower potential offsets, see e.g. Ref. \citenum{C5EE00751H}. This coincides with the consensus that the pristine basal plane is not the main origin of \ce{MoS2} activity, and edges and defects must be considered. 
When sites on certain facets are more active than others, the experimentally observed activity will depend largely on the morphology of prepared samples. Also the orientation of \ce{MoS2} crystals with respect to the electrode substrate is important, as there is considerable resistance associated with electron transport between layers \cite{doi:10.1021/nl403620g}.

In the work by Humphrey {\it et al.} \cite{C9NR10702A}, planar support is thought to have produced relatively low edge-content \ce{MoS2} for the pristine case and with low levels of \ce{Co}-doping. As noted there, the larger overpotential compared to other studies suggests that this activity is more representative of the basal plane. This inherent activity of the basal plane cannot be explained without introducing defects, as the pristine adsorption sites are much too unfavorable for evolution to occur. Most synthesized \ce{MoS2} does however contain a significant number of sulfur vacancies, with certain deposition methods yielding stoichiometries of \ce{MoS_{1.6}} \cite{C4NR04317K} and \ce{MoS_{1.8}} \cite{doi:10.1021/acscatal.6b01274}, although this total deficiency will also depend in part 
on
the stoichiometry and prevalence of the edge terminations. These vacancies create local sites with adsorption energies of roughly $0.1\,\mathrm{eV}$. However, neighboring sites are not significantly activated and (assuming evenly dispersed \ce{S}-vacancies) the reaction would still be limited to a Volmer-Heyrovsky mechanism. More complex defect configurations could possibly facilitate the Tafel mechanism. For reference, the Tafel slope of the basal plane of pristine \ce{MoS2} under acidic conditions has been measured to be around $120$~mV~dec$^{-1}$ \cite{C7SC02545A}, indicative (but not conclusive) of a mechanism rate-determined by the Volmer step, from which neither the Tafel nor Heyrovsky path can be disregarded. 

Importantly, the aforementioned study finds that low levels of \ce{Co}-doping is detrimental to the inherent activity, in contradiction to the results of the present study in which only stoichiometric \ce{MoS2} is considered. This indicates that the combination of atomic doping and intrinsic defects can lead to overall deactivation of the basal plane. In Ref. \citenum{C9NR10702A}, this is supported by the calculated adsorption energies and equilibrium defect configurations, and it would be 
interesting to study this problem in terms of activation barriers and possible reaction paths of the \ce{Co}- and \ce{Ni}-doped defect systems. 

We summarize that a direct comparison of our results with experiment is not possible without also considering \ce{MoS2} edge sites and defects in the basal plane, as well as other possible doping configurations and interplay with inherent sulfur vacancies. A systematic investigation of activation barriers and coverage dependence across these systems is needed for drawing solid conclusions.

The overall trend of relative activity is clear in Figure \ref{fig:barrs}b, the optimum lies in the vicinity of $\Delta \Omega_\mathrm{H}\approx 0$~eV, though we note that the peaks of the directly fitted volcanoes are in this case centered at $0.3$~eV. The discrepancy between the solid line and the points illustrates the expected inaccuracies associated with $\Delta G_\mathrm{H}$ as a descriptor, as the relation to the activation energy is only implicit. Overall the description works well, but comparing individual systems in terms of only $\Delta G_\mathrm{H}$ will be prone to errors.
On a larger scale, it is important to note that the condition $\Delta G_\mathrm{H}\approx 0$~eV is not sufficient to guarantee high absolute HER activity, but is merely a necessary condition to be near the optimum for a given class of systems. 
As an example, we see in this study that the Heyrovsky mechanism on local impurity-induced \ce{S}-sites in \ce{MoS2} behaves markedly differently from {\it e.g.} the Tafel mechanism on uniformly covered \ce{Pt}(111), even if the \ce{H} adsorption energy is near zero in both cases. 

\section{Conclusions}
Methodology for computing reaction and activation energies under electrochemical conditions 
corresponding to a specified applied voltage was used to investigate the HER mechanism on the basal plane of 2H-\ce{MoS2} with single-atom 3d-metal doping. The effect of the aqueous solvent is included by using a cluster of a four H$_2$O molecules and a proton (Eigen cation) in the neighborhood of the active site and then a polarizable continuum solvent for the rest of the solvent phase. The effect of the various transition-metal dopants spans a wide range of activation energies and within this scope \ce{Ni} stands out with the best overall activity at moderate negative applied potential. The reaction barriers can be correlated with the potential-dependent promotion energy $E_\text{pd}$, and the modelled kinetics has the characteristic volcano-shape with respect to the hydrogen adsorption energy. The results were compared with the commonly used $\Delta G_\mathrm{H}$ model, and the condition $\Delta G_\mathrm{H}\approx 0$~eV was shown not to be sufficient for predicting a low reaction barrier in these systems. 

The calculations used a single monolayer of 2H-\ce{MoS2}, but the results are expected to be representative for multi-layer slabs as the adsorption energies are found to be similar. 
The reaction was found to proceed predominantly through a Volmer-Heyrovsky path, where the dopant-activated sulfur sites provided a large reduction in the adsorption energy, and therefore, also in the activation energy of the Volmer step which on pristine \ce{MoS2} is so unfavorable that it essentially does not occur. Thus, the inherent activity of \ce{MoS2} observed in experiments must come from defect sites (e.g. sulfur vacancies on the basal plane) and/or from edge sites. These sites are expected to be more active than the pristine basal plane, and a full picture of the various possible sites is necessary for a more in-depth comparison with experimental data. Such an investigation is planned for future work.

\section*{Conflicts of interest}
There are no conflicts to declare.

\section*{Acknowledgements}
We thank K. Laasonen for discussions. The 
calculations were performed on resources provided by Sigma2 - the National Infrastructure for High Performance Computing and Data Storage in Norway, project No. NN9497K. J. A. acknowledges financial support from the Academy of Finland, project No. 322832 ‘‘NANOIONICS’’. 
H. J. acknowledges financial support from the Icelandic Research Fundproject No. 207283-053.



\balance


\bibliography{bi} 

\providecommand*{\mcitethebibliography}{\thebibliography}
\csname @ifundefined\endcsname{endmcitethebibliography}
{\let\endmcitethebibliography\endthebibliography}{}
\begin{mcitethebibliography}{43}
\providecommand*{\natexlab}[1]{#1}
\providecommand*{\mciteSetBstSublistMode}[1]{}
\providecommand*{\mciteSetBstMaxWidthForm}[2]{}
\providecommand*{\mciteBstWouldAddEndPuncttrue}
  {\def\EndOfBibitem{\unskip.}}
\providecommand*{\mciteBstWouldAddEndPunctfalse}
  {\let\EndOfBibitem\relax}
\providecommand*{\mciteSetBstMidEndSepPunct}[3]{}
\providecommand*{\mciteSetBstSublistLabelBeginEnd}[3]{}
\providecommand*{\EndOfBibitem}{}
\mciteSetBstSublistMode{f}
\mciteSetBstMaxWidthForm{subitem}
{(\emph{\alph{mcitesubitemcount}})}
\mciteSetBstSublistLabelBeginEnd{\mcitemaxwidthsubitemform\space}
{\relax}{\relax}

\bibitem[Dawood \emph{et~al.}(2020)Dawood, Anda, and
  Shafiullah]{DAWOOD20203847}
F.~Dawood, M.~Anda and G.~Shafiullah, \emph{International Journal of Hydrogen
  Energy}, 2020, \textbf{45}, 3847--3869\relax
\mciteBstWouldAddEndPuncttrue
\mciteSetBstMidEndSepPunct{\mcitedefaultmidpunct}
{\mcitedefaultendpunct}{\mcitedefaultseppunct}\relax
\EndOfBibitem
\bibitem[Holladay \emph{et~al.}(2009)Holladay, Hu, King, and
  Wang]{HOLLADAY2009244}
J.~Holladay, J.~Hu, D.~King and Y.~Wang, \emph{Catalysis Today}, 2009,
  \textbf{139}, 244--260\relax
\mciteBstWouldAddEndPuncttrue
\mciteSetBstMidEndSepPunct{\mcitedefaultmidpunct}
{\mcitedefaultendpunct}{\mcitedefaultseppunct}\relax
\EndOfBibitem
\bibitem[Faber and Jin(2014)]{C4EE01760A}
M.~S. Faber and S.~Jin, \emph{Energy Environ. Sci.}, 2014, \textbf{7},
  3519--3542\relax
\mciteBstWouldAddEndPuncttrue
\mciteSetBstMidEndSepPunct{\mcitedefaultmidpunct}
{\mcitedefaultendpunct}{\mcitedefaultseppunct}\relax
\EndOfBibitem
\bibitem[Hinnemann \emph{et~al.}(2005)Hinnemann, Moses, Bonde, Jørgensen,
  Nielsen, Horch, Chorkendorff, and Nørskov]{doi:10.1021/ja0504690}
B.~Hinnemann, P.~G. Moses, J.~Bonde, K.~P. Jørgensen, J.~H. Nielsen, S.~Horch,
  I.~Chorkendorff and J.~K. Nørskov, \emph{Journal of the American Chemical
  Society}, 2005, \textbf{127}, 5308--5309\relax
\mciteBstWouldAddEndPuncttrue
\mciteSetBstMidEndSepPunct{\mcitedefaultmidpunct}
{\mcitedefaultendpunct}{\mcitedefaultseppunct}\relax
\EndOfBibitem
\bibitem[Bonde \emph{et~al.}(2009)Bonde, Moses, Jaramillo, Nørskov, and
  Chorkendorff]{B803857K}
J.~Bonde, P.~G. Moses, T.~F. Jaramillo, J.~K. Nørskov and I.~Chorkendorff,
  \emph{Faraday Discuss.}, 2009, \textbf{140}, 219--231\relax
\mciteBstWouldAddEndPuncttrue
\mciteSetBstMidEndSepPunct{\mcitedefaultmidpunct}
{\mcitedefaultendpunct}{\mcitedefaultseppunct}\relax
\EndOfBibitem
\bibitem[Benck \emph{et~al.}(2012)Benck, Chen, Kuritzky, Forman, and
  Jaramillo]{doi:10.1021/cs300451q}
J.~D. Benck, Z.~Chen, L.~Y. Kuritzky, A.~J. Forman and T.~F. Jaramillo,
  \emph{ACS Catalysis}, 2012, \textbf{2}, 1916--1923\relax
\mciteBstWouldAddEndPuncttrue
\mciteSetBstMidEndSepPunct{\mcitedefaultmidpunct}
{\mcitedefaultendpunct}{\mcitedefaultseppunct}\relax
\EndOfBibitem
\bibitem[Kibsgaard \emph{et~al.}(2012)Kibsgaard, Chen, Reinecke, and
  Jaramillo]{Kibsgaard2012}
J.~Kibsgaard, Z.~Chen, B.~N. Reinecke and T.~F. Jaramillo, \emph{Nature
  Materials}, 2012, \textbf{11}, 963--969\relax
\mciteBstWouldAddEndPuncttrue
\mciteSetBstMidEndSepPunct{\mcitedefaultmidpunct}
{\mcitedefaultendpunct}{\mcitedefaultseppunct}\relax
\EndOfBibitem
\bibitem[Jaramillo \emph{et~al.}(2007)Jaramillo, Jørgensen, Bonde, Nielsen,
  Horch, and Chorkendorff]{doi:10.1126/science.1141483}
T.~F. Jaramillo, K.~P. Jørgensen, J.~Bonde, J.~H. Nielsen, S.~Horch and
  I.~Chorkendorff, \emph{Science}, 2007, \textbf{317}, 100--102\relax
\mciteBstWouldAddEndPuncttrue
\mciteSetBstMidEndSepPunct{\mcitedefaultmidpunct}
{\mcitedefaultendpunct}{\mcitedefaultseppunct}\relax
\EndOfBibitem
\bibitem[Kong \emph{et~al.}(2013)Kong, Wang, Cha, Pasta, Koski, Yao, and
  Cui]{doi:10.1021/nl400258t}
D.~Kong, H.~Wang, J.~J. Cha, M.~Pasta, K.~J. Koski, J.~Yao and Y.~Cui,
  \emph{Nano Letters}, 2013, \textbf{13}, 1341--1347\relax
\mciteBstWouldAddEndPuncttrue
\mciteSetBstMidEndSepPunct{\mcitedefaultmidpunct}
{\mcitedefaultendpunct}{\mcitedefaultseppunct}\relax
\EndOfBibitem
\bibitem[Li \emph{et~al.}(2016)Li, Tsai, Koh, Cai, Contryman, Fragapane, Zhao,
  Han, Manoharan, Abild-Pedersen, N{\o}rskov, and Zheng]{Li2016}
H.~Li, C.~Tsai, A.~L. Koh, L.~Cai, A.~W. Contryman, A.~H. Fragapane, J.~Zhao,
  H.~S. Han, H.~C. Manoharan, F.~Abild-Pedersen, J.~K. N{\o}rskov and X.~Zheng,
  \emph{Nature Materials}, 2016, \textbf{15}, 48--53\relax
\mciteBstWouldAddEndPuncttrue
\mciteSetBstMidEndSepPunct{\mcitedefaultmidpunct}
{\mcitedefaultendpunct}{\mcitedefaultseppunct}\relax
\EndOfBibitem
\bibitem[Zhang \emph{et~al.}(2020)Zhang, Zhu, Guo, Cao, Wu, Wang, and
  Lu]{C9CY01901D}
T.~Zhang, H.~Zhu, C.~Guo, S.~Cao, C.-M.~L. Wu, Z.~Wang and X.~Lu, \emph{Catal.
  Sci. Technol.}, 2020, \textbf{10}, 458--465\relax
\mciteBstWouldAddEndPuncttrue
\mciteSetBstMidEndSepPunct{\mcitedefaultmidpunct}
{\mcitedefaultendpunct}{\mcitedefaultseppunct}\relax
\EndOfBibitem
\bibitem[Deng \emph{et~al.}(2015)Deng, Li, Xiao, Tu, Deng, Yang, Tian, Li, Ren,
  and Bao]{C5EE00751H}
J.~Deng, H.~Li, J.~Xiao, Y.~Tu, D.~Deng, H.~Yang, H.~Tian, J.~Li, P.~Ren and
  X.~Bao, \emph{Energy Environ. Sci.}, 2015, \textbf{8}, 1594--1601\relax
\mciteBstWouldAddEndPuncttrue
\mciteSetBstMidEndSepPunct{\mcitedefaultmidpunct}
{\mcitedefaultendpunct}{\mcitedefaultseppunct}\relax
\EndOfBibitem
\bibitem[Wang \emph{et~al.}(2015)Wang, Tsai, Kong, Chan, Abild-Pedersen,
  N{\o}rskov, and Cui]{Wang2015}
H.~Wang, C.~Tsai, D.~Kong, K.~Chan, F.~Abild-Pedersen, J.~K. N{\o}rskov and
  Y.~Cui, \emph{Nano Research}, 2015, \textbf{8}, 566--575\relax
\mciteBstWouldAddEndPuncttrue
\mciteSetBstMidEndSepPunct{\mcitedefaultmidpunct}
{\mcitedefaultendpunct}{\mcitedefaultseppunct}\relax
\EndOfBibitem
\bibitem[Merki \emph{et~al.}(2012)Merki, Vrubel, Rovelli, Fierro, and
  Hu]{C2SC20539D}
D.~Merki, H.~Vrubel, L.~Rovelli, S.~Fierro and X.~Hu, \emph{Chem. Sci.}, 2012,
  \textbf{3}, 2515--2525\relax
\mciteBstWouldAddEndPuncttrue
\mciteSetBstMidEndSepPunct{\mcitedefaultmidpunct}
{\mcitedefaultendpunct}{\mcitedefaultseppunct}\relax
\EndOfBibitem
\bibitem[Escalera-López \emph{et~al.}(2016)Escalera-López, Niu, Yin, Cooke,
  Rees, and Palmer]{doi:10.1021/acscatal.6b01274}
D.~Escalera-López, Y.~Niu, J.~Yin, K.~Cooke, N.~V. Rees and R.~E. Palmer,
  \emph{ACS Catalysis}, 2016, \textbf{6}, 6008--6017\relax
\mciteBstWouldAddEndPuncttrue
\mciteSetBstMidEndSepPunct{\mcitedefaultmidpunct}
{\mcitedefaultendpunct}{\mcitedefaultseppunct}\relax
\EndOfBibitem
\bibitem[Lau \emph{et~al.}(2018)Lau, Lu, Kulhavý, Wu, Lu, Wu, Kato, Foord,
  Soo, Suenaga, and Tsang]{C8SC01114A}
T.~H.~M. Lau, X.~Lu, J.~Kulhavý, S.~Wu, L.~Lu, T.-S. Wu, R.~Kato, J.~S. Foord,
  Y.-L. Soo, K.~Suenaga and S.~C.~E. Tsang, \emph{Chem. Sci.}, 2018,
  \textbf{9}, 4769--4776\relax
\mciteBstWouldAddEndPuncttrue
\mciteSetBstMidEndSepPunct{\mcitedefaultmidpunct}
{\mcitedefaultendpunct}{\mcitedefaultseppunct}\relax
\EndOfBibitem
\bibitem[Humphrey \emph{et~al.}(2020)Humphrey, Kronberg, Cai, Laasonen, Palmer,
  and Wain]{C9NR10702A}
J.~J.~L. Humphrey, R.~Kronberg, R.~Cai, K.~Laasonen, R.~E. Palmer and A.~J.
  Wain, \emph{Nanoscale}, 2020, \textbf{12}, 4459--4472\relax
\mciteBstWouldAddEndPuncttrue
\mciteSetBstMidEndSepPunct{\mcitedefaultmidpunct}
{\mcitedefaultendpunct}{\mcitedefaultseppunct}\relax
\EndOfBibitem
\bibitem[Hakala \emph{et~al.}(2017)Hakala, Kronberg, and Laasonen]{Hakala2017}
M.~Hakala, R.~Kronberg and K.~Laasonen, \emph{Scientific Reports}, 2017,
  \textbf{7}, 15243\relax
\mciteBstWouldAddEndPuncttrue
\mciteSetBstMidEndSepPunct{\mcitedefaultmidpunct}
{\mcitedefaultendpunct}{\mcitedefaultseppunct}\relax
\EndOfBibitem
\bibitem[Hammer \emph{et~al.}(1999)Hammer, Hansen, and
  N\o{}rskov]{PhysRevB.59.7413}
B.~Hammer, L.~B. Hansen and J.~K. N\o{}rskov, \emph{Phys. Rev. B}, 1999,
  \textbf{59}, 7413--7421\relax
\mciteBstWouldAddEndPuncttrue
\mciteSetBstMidEndSepPunct{\mcitedefaultmidpunct}
{\mcitedefaultendpunct}{\mcitedefaultseppunct}\relax
\EndOfBibitem
\bibitem[Grimme \emph{et~al.}(2010)Grimme, Antony, Ehrlich, and
  Krieg]{doi:10.1063/1.3382344}
S.~Grimme, J.~Antony, S.~Ehrlich and H.~Krieg, \emph{The Journal of Chemical
  Physics}, 2010, \textbf{132}, 154104\relax
\mciteBstWouldAddEndPuncttrue
\mciteSetBstMidEndSepPunct{\mcitedefaultmidpunct}
{\mcitedefaultendpunct}{\mcitedefaultseppunct}\relax
\EndOfBibitem
\bibitem[Bl\"ochl(1994)]{PhysRevB.50.17953}
P.~E. Bl\"ochl, \emph{Phys. Rev. B}, 1994, \textbf{50}, 17953--17979\relax
\mciteBstWouldAddEndPuncttrue
\mciteSetBstMidEndSepPunct{\mcitedefaultmidpunct}
{\mcitedefaultendpunct}{\mcitedefaultseppunct}\relax
\EndOfBibitem
\bibitem[Himmetoglu \emph{et~al.}(2014)Himmetoglu, Floris, de~Gironcoli, and
  Cococcioni]{doi.org/10.1002/qua.24521}
B.~Himmetoglu, A.~Floris, S.~de~Gironcoli and M.~Cococcioni,
  \emph{International Journal of Quantum Chemistry}, 2014, \textbf{114},
  14--49\relax
\mciteBstWouldAddEndPuncttrue
\mciteSetBstMidEndSepPunct{\mcitedefaultmidpunct}
{\mcitedefaultendpunct}{\mcitedefaultseppunct}\relax
\EndOfBibitem
\bibitem[Perdew \emph{et~al.}(1996)Perdew, Ernzerhof, and
  Burke]{doi:10.1063/1.472933}
J.~P. Perdew, M.~Ernzerhof and K.~Burke, \emph{The Journal of Chemical
  Physics}, 1996, \textbf{105}, 9982--9985\relax
\mciteBstWouldAddEndPuncttrue
\mciteSetBstMidEndSepPunct{\mcitedefaultmidpunct}
{\mcitedefaultendpunct}{\mcitedefaultseppunct}\relax
\EndOfBibitem
\bibitem[Adamo and Barone(1999)]{doi:10.1063/1.478522}
C.~Adamo and V.~Barone, \emph{The Journal of Chemical Physics}, 1999,
  \textbf{110}, 6158--6170\relax
\mciteBstWouldAddEndPuncttrue
\mciteSetBstMidEndSepPunct{\mcitedefaultmidpunct}
{\mcitedefaultendpunct}{\mcitedefaultseppunct}\relax
\EndOfBibitem
\bibitem[Kresse and Joubert(1999)]{PhysRevB.59.1758}
G.~Kresse and D.~Joubert, \emph{Phys. Rev. B}, 1999, \textbf{59},
  1758--1775\relax
\mciteBstWouldAddEndPuncttrue
\mciteSetBstMidEndSepPunct{\mcitedefaultmidpunct}
{\mcitedefaultendpunct}{\mcitedefaultseppunct}\relax
\EndOfBibitem
\bibitem[Henkelman \emph{et~al.}(2000)Henkelman, Uberuaga, and
  Jónsson]{doi:10.1063/1.1329672}
G.~Henkelman, B.~P. Uberuaga and H.~Jónsson, \emph{The Journal of Chemical
  Physics}, 2000, \textbf{113}, 9901--9904\relax
\mciteBstWouldAddEndPuncttrue
\mciteSetBstMidEndSepPunct{\mcitedefaultmidpunct}
{\mcitedefaultendpunct}{\mcitedefaultseppunct}\relax
\EndOfBibitem
\bibitem[Henkelman and Jónsson(1999)]{doi:10.1063/1.480097}
G.~Henkelman and H.~Jónsson, \emph{The Journal of Chemical Physics}, 1999,
  \textbf{111}, 7010--7022\relax
\mciteBstWouldAddEndPuncttrue
\mciteSetBstMidEndSepPunct{\mcitedefaultmidpunct}
{\mcitedefaultendpunct}{\mcitedefaultseppunct}\relax
\EndOfBibitem
\bibitem[Trasatti(1972)]{TRASATTI1972163}
S.~Trasatti, \emph{Journal of Electroanalytical Chemistry and Interfacial
  Electrochemistry}, 1972, \textbf{39}, 163--184\relax
\mciteBstWouldAddEndPuncttrue
\mciteSetBstMidEndSepPunct{\mcitedefaultmidpunct}
{\mcitedefaultendpunct}{\mcitedefaultseppunct}\relax
\EndOfBibitem
\bibitem[N{\o}rskov \emph{et~al.}(2005)N{\o}rskov, Bligaard, Logadottir,
  Kitchin, Chen, Pandelov, and Stimming]{N_rskov_2005}
J.~K. N{\o}rskov, T.~Bligaard, A.~Logadottir, J.~R. Kitchin, J.~G. Chen,
  S.~Pandelov and U.~Stimming, \emph{Journal of The Electrochemical Society},
  2005, \textbf{152}, J23\relax
\mciteBstWouldAddEndPuncttrue
\mciteSetBstMidEndSepPunct{\mcitedefaultmidpunct}
{\mcitedefaultendpunct}{\mcitedefaultseppunct}\relax
\EndOfBibitem
\bibitem[Nørskov \emph{et~al.}(2004)Nørskov, Rossmeisl, Logadottir,
  Lindqvist, Kitchin, Bligaard, and Jónsson]{doi:10.1021/jp047349j}
J.~K. Nørskov, J.~Rossmeisl, A.~Logadottir, L.~Lindqvist, J.~R. Kitchin,
  T.~Bligaard and H.~Jónsson, \emph{The Journal of Physical Chemistry B},
  2004, \textbf{108}, 17886--17892\relax
\mciteBstWouldAddEndPuncttrue
\mciteSetBstMidEndSepPunct{\mcitedefaultmidpunct}
{\mcitedefaultendpunct}{\mcitedefaultseppunct}\relax
\EndOfBibitem
\bibitem[Mathew \emph{et~al.}(2014)Mathew, Sundararaman, Letchworth-Weaver,
  Arias, and Hennig]{doi:10.1063/1.4865107}
K.~Mathew, R.~Sundararaman, K.~Letchworth-Weaver, T.~A. Arias and R.~G. Hennig,
  \emph{The Journal of Chemical Physics}, 2014, \textbf{140}, 084106\relax
\mciteBstWouldAddEndPuncttrue
\mciteSetBstMidEndSepPunct{\mcitedefaultmidpunct}
{\mcitedefaultendpunct}{\mcitedefaultseppunct}\relax
\EndOfBibitem
\bibitem[Mathew \emph{et~al.}(2019)Mathew, Kolluru, Mula, Steinmann, and
  Hennig]{doi:10.1063/1.5132354}
K.~Mathew, V.~S.~C. Kolluru, S.~Mula, S.~N. Steinmann and R.~G. Hennig,
  \emph{The Journal of Chemical Physics}, 2019, \textbf{151}, 234101\relax
\mciteBstWouldAddEndPuncttrue
\mciteSetBstMidEndSepPunct{\mcitedefaultmidpunct}
{\mcitedefaultendpunct}{\mcitedefaultseppunct}\relax
\EndOfBibitem
\bibitem[Rossmeisl \emph{et~al.}(2008)Rossmeisl, Skúlason, Björketun,
  Tripkovic, and Nørskov]{ROSSMEISL200868}
J.~Rossmeisl, E.~Skúlason, M.~E. Björketun, V.~Tripkovic and J.~K. Nørskov,
  \emph{Chemical Physics Letters}, 2008, \textbf{466}, 68--71\relax
\mciteBstWouldAddEndPuncttrue
\mciteSetBstMidEndSepPunct{\mcitedefaultmidpunct}
{\mcitedefaultendpunct}{\mcitedefaultseppunct}\relax
\EndOfBibitem
\bibitem[Van~den Bossche \emph{et~al.}(2019)Van~den Bossche, Skúlason,
  Rose-Petruck, and Jónsson]{doi:10.1021/acs.jpcc.8b10046}
M.~Van~den Bossche, E.~Skúlason, C.~Rose-Petruck and H.~Jónsson, \emph{The
  Journal of Physical Chemistry C}, 2019, \textbf{123}, 4116--4124\relax
\mciteBstWouldAddEndPuncttrue
\mciteSetBstMidEndSepPunct{\mcitedefaultmidpunct}
{\mcitedefaultendpunct}{\mcitedefaultseppunct}\relax
\EndOfBibitem
\bibitem[198(1986)]{1986417}
\emph{Journal of Electroanalytical Chemistry and Interfacial Electrochemistry},
  1986, \textbf{209}, 417--428\relax
\mciteBstWouldAddEndPuncttrue
\mciteSetBstMidEndSepPunct{\mcitedefaultmidpunct}
{\mcitedefaultendpunct}{\mcitedefaultseppunct}\relax
\EndOfBibitem
\bibitem[Santos and Schmickler(2004)]{SANTOS200426}
E.~Santos and W.~Schmickler, \emph{Chemical Physics Letters}, 2004,
  \textbf{400}, 26--29\relax
\mciteBstWouldAddEndPuncttrue
\mciteSetBstMidEndSepPunct{\mcitedefaultmidpunct}
{\mcitedefaultendpunct}{\mcitedefaultseppunct}\relax
\EndOfBibitem
\bibitem[Wakabayashi \emph{et~al.}(1975)Wakabayashi, Smith, and
  Nicklow]{PhysRevB.12.659}
N.~Wakabayashi, H.~G. Smith and R.~M. Nicklow, \emph{Phys. Rev. B}, 1975,
  \textbf{12}, 659--663\relax
\mciteBstWouldAddEndPuncttrue
\mciteSetBstMidEndSepPunct{\mcitedefaultmidpunct}
{\mcitedefaultendpunct}{\mcitedefaultseppunct}\relax
\EndOfBibitem
\bibitem[B\"oker \emph{et~al.}(2001)B\"oker, Severin, M\"uller, Janowitz,
  Manzke, Vo\ss{}, Kr\"uger, Mazur, and Pollmann]{PhysRevB.64.235305}
T.~B\"oker, R.~Severin, A.~M\"uller, C.~Janowitz, R.~Manzke, D.~Vo\ss{},
  P.~Kr\"uger, A.~Mazur and J.~Pollmann, \emph{Phys. Rev. B}, 2001,
  \textbf{64}, 235305\relax
\mciteBstWouldAddEndPuncttrue
\mciteSetBstMidEndSepPunct{\mcitedefaultmidpunct}
{\mcitedefaultendpunct}{\mcitedefaultseppunct}\relax
\EndOfBibitem
\bibitem[Mak \emph{et~al.}(2010)Mak, Lee, Hone, Shan, and
  Heinz]{PhysRevLett.105.136805}
K.~F. Mak, C.~Lee, J.~Hone, J.~Shan and T.~F. Heinz, \emph{Phys. Rev. Lett.},
  2010, \textbf{105}, 136805\relax
\mciteBstWouldAddEndPuncttrue
\mciteSetBstMidEndSepPunct{\mcitedefaultmidpunct}
{\mcitedefaultendpunct}{\mcitedefaultseppunct}\relax
\EndOfBibitem
\bibitem[Liu \emph{et~al.}(2020)Liu, Hybertsen, and
  Wu]{https://doi.org/10.1002/anie.202003091}
M.~Liu, M.~S. Hybertsen and Q.~Wu, \emph{Angewandte Chemie International
  Edition}, 2020, \textbf{59}, 14835--14841\relax
\mciteBstWouldAddEndPuncttrue
\mciteSetBstMidEndSepPunct{\mcitedefaultmidpunct}
{\mcitedefaultendpunct}{\mcitedefaultseppunct}\relax
\EndOfBibitem
\bibitem[Yu \emph{et~al.}(2014)Yu, Huang, Li, Steinmann, Yang, and
  Cao]{doi:10.1021/nl403620g}
Y.~Yu, S.-Y. Huang, Y.~Li, S.~N. Steinmann, W.~Yang and L.~Cao, \emph{Nano
  Letters}, 2014, \textbf{14}, 553--558\relax
\mciteBstWouldAddEndPuncttrue
\mciteSetBstMidEndSepPunct{\mcitedefaultmidpunct}
{\mcitedefaultendpunct}{\mcitedefaultseppunct}\relax
\EndOfBibitem
\bibitem[Cuddy \emph{et~al.}(2014)Cuddy, Arkill, Wang, Komsa, Krasheninnikov,
  and Palmer]{C4NR04317K}
M.~J. Cuddy, K.~P. Arkill, Z.~W. Wang, H.-P. Komsa, A.~V. Krasheninnikov and
  R.~E. Palmer, \emph{Nanoscale}, 2014, \textbf{6}, 12463--12469\relax
\mciteBstWouldAddEndPuncttrue
\mciteSetBstMidEndSepPunct{\mcitedefaultmidpunct}
{\mcitedefaultendpunct}{\mcitedefaultseppunct}\relax
\EndOfBibitem
\bibitem[Bentley \emph{et~al.}(2017)Bentley, Kang, Maddar, Li, Walker, Zhang,
  and Unwin]{C7SC02545A}
C.~L. Bentley, M.~Kang, F.~M. Maddar, F.~Li, M.~Walker, J.~Zhang and P.~R.
  Unwin, \emph{Chem. Sci.}, 2017, \textbf{8}, 6583--6593\relax
\mciteBstWouldAddEndPuncttrue
\mciteSetBstMidEndSepPunct{\mcitedefaultmidpunct}
{\mcitedefaultendpunct}{\mcitedefaultseppunct}\relax
\EndOfBibitem
\end{mcitethebibliography}
\bibliographystyle{rsc} 

\end{document}